\documentclass[%
 reprint,
 amsmath,amssymb,
 aps,
pra,
]{revtex4-2}
\usepackage{amsthm,color}
\newtheorem*{postulate}{Postulate}
\DeclareMathOperator{\Rea}{Re}
\DeclareMathOperator{\Ima}{Im}

\usepackage{graphicx}
\usepackage{bm}
\usepackage{hyperref}
\usepackage{esint}

\begin{document}
\title{Generator of spatial evolution of the electromagnetic field}%

\author{Dmitri B. Horoshko}\email{horoshko@ifanbel.bas-net.by}
\affiliation{B. I. Stepanov Institute of Physics, NASB, Nezavisimosti Ave 68, Minsk 220072 Belarus}
\date{\today}

\begin{abstract}
Starting with Maxwell's equations and defining normal variables in the Fourier space, we write the equations of temporal evolution of the electromagnetic field with sources in the Hamiltonian and Lagrangian forms, making explicit all intermediate steps often omitted in standard textbooks. Then, we follow the same steps to write the equations of evolution of this field along a spatial dimension in the Hamiltonian and Lagrangian forms. In this way, we arrive at the explicit form of the generator of spatial evolution of the electromagnetic field with sources and show that it has a physical meaning of the modulus of momentum transferred through a given plane orthogonal to the direction of propagation. In a particular case of free field this generator coincides with the projection of the full momentum of the field on the propagation direction, taken with a negative sign. The Hamiltonian and Lagrangian formulations of the spatial evolution are indispensable for a correct quantization of the field when considering its spatial rather than temporal evolution, in particular, for a correct definition of the equal-space commutation relations.
\end{abstract}
\maketitle

\section{Introduction \label{sec:intro}}
The evolution of the electromagnetic field is described classically by Maxwell's equations which can be written in the Lagrangian or Hamiltonian form \cite{Cohen-Tannoudji89}. Canonical field quantization is performed by transforming the Poisson bracket into the equal-time commutator for the generalized coordinates and momenta, while the Hamiltonian operator plays the role of the generator of the temporal evolution. The Heisenberg and Schr\"odinger equations allow one to calculate the state of the system at any moment of time $t$, given its state is known at time $t_0$. This formalism is well suited for the description of temporal evolution of a field in a cavity, especially when the field is decomposed into a sum of spatial modes, the amplitudes of which obey a set of coupled temporal differential equations.

With the appearance of nonlinear optics, however, a different approach was adopted for the description of single passage of radiation through a nonlinear medium. In this approach, a spatial evolution of the field envelope is considered along the nonlinear medium length \cite{Bloembergen65} and a spatial differential equation is deduced, allowing one to calculate the field at point $z$ provided it was known at point $z_0$. It was proposed by Shen \cite{Shen67} to consider this equation for a quantized field as a spatial analog of the Heisenberg equation, where the role of the generator of spatial evolution is played by the component of the field momentum along the direction of propagation. This approach was further developed by Caves and Crouch, who applied it to the description of an optical parametric amplifier \cite{Caves87}. Decomposing the field into a sum of temporal modes and postulating the equal-space commutation relations for their quantum amplitudes, they arrived at a set of coupled spatial differential equations for the modal amplitudes. This approach proved to be highly efficient for the description of spatial field evolution in nonlinear media \cite{Kolobov99} and is widely employed at present for this purpose, including generation of entangled beams from a monochromatic pump in counterpropagating geometry \cite{Corti16} or in aperiodically poled crystals \cite{Horoshko17}, as well as for generation of pulsed squeezed light in bulk crystals \cite{Wasilewski06} and waveguides \cite{Quesada20}.  

The exact form of the spatial evolution generator and its physical meaning were, however, for a long time obscure. The existence of this generator is implicit in Shen's formalism, but its exact form is not necessary for practical calculations in the Heisenberg picture, since the differential equations are obtained directly from Maxwell's equations for the quantized field. Establishing this generator is of paramount importance though, first, because it is a fundamental problem of quantum electrodynamics important for the correct field quantization, and, second, because this generator is required for practical calculations in the Schr\"odinger picture. For example, finding the modes of squeezing of an optical parametric amplifier requires a diagonalization of the squeezing matrix, given by a spatial integral of this generator \cite{Horoshko19,LaVolpe21}. In the last years, the spatial evolution generator has acquired special importance for the description of arrays of coupled nonlinear waveguides, playing the role of the ``discrete lattice Hamiltonian'' \cite{Smirnova20} and helping to find the lattice supermodes \cite{Barral20a,Barral20b}, which may be topologically protected under certain conditions \cite{Blanco18}. In addition, modeling the generation of highly squeezed quantum fields requires calculation of the time \cite{Christ13,Quesada15} or space-ordering \cite{Lipfert18} terms of the evolution operator. The latter are defined by the commutator of the spatial generator with itself at different positions. Thus, the knowledge of the exact expression for the generator of spatial evolution is highly desirable for practical calculations. 

Abram considered a spatially one-dimensional field propagating in both directions in a dispersionless linear dielectric and showed that the spatial evolution generator, under some simplifying assumptions, is proportional to the difference of the energy densities of the forward and backward propagating waves \cite{Abram87}. Ben-Aryeh and coworkers considered a possibility to postulate the spatial generator as an integral of the momentum density of the field in the spatially one-dimensional \cite{Huttner90} and three-dimensional \cite{Serulnik91} cases. Later, Ben-Aryeh and Serulnik  discussed difficulties of this approach and suggested another form for the spatial generator: a three-dimensional integral of the momentum flux density over the transverse plane and time \cite{Ben-Aryeh91}. This approach showed excellent results for propagation in one direction, but met difficulties in describing a two-directional propagation \cite{Toren94}, which invoked various \textit{ad hoc} corrections, including the exchange of the roles of the photon creation and annihilation operators for the backward-propagating field \cite{Ben-Aryeh92} or inverting the sign of the spatial derivative for this field \cite{Perina95,Perina00}. Various approaches to the description of light propagation were compared and discussed in a review \cite{Luks02} and a book \cite{Luks09}. 

The aim of the present paper is to construct the spatial evolution generator of the fully four-dimensional electromagnetic field with sources directly from Maxwell's equations without any approximations and simplifications. For this purpose we employ an approach which can be called ``inductive'' because it consists in starting from the equations of motion for particular fields, obtained in their turn by an inductive generalization of experimental observations, and proceeding to more general expressions for the system Hamiltonian and then Lagrangian. Induction, in general, can be understood as ``an inference from particular objects, phenomena to a common conclusion, from separate facts to their generalizations'' \cite{Novikov-Novikov}. The reciprocal notion is deduction, which is ``an inference from the common to the particular, from general judgements to particular conclusions'' \cite{Novikov-Novikov}. The traditional approach to electrodynamics can thus be called ``deductive'': it starts with the most general principle -- the principle of least action, where the action is defined as a four-dimensional integral of a postulated Lagrangian density, and then the Hamiltonian density and the equations of motion for particular fields are deduced from the latter \cite{Dirac64,Akhiezer65}. Our inductive approach moves in the opposite direction and could, in principle, end up with a formulation of a spatial counterpart of the least action principle, which we postpone for the future.

In order to make our approach clear, in Sec.~\ref{sec:Time}, we recast the traditional consideration of temporal evolution of the electromagnetic field with sources in the inductive way, obtaining the field Hamiltonian, given in this case by the field energy, and the standard Lagrangian of the electromagnetic field from Maxwell's equations. Our main achievement is, however, not giving these well-known expressions, but giving a clear sequence of steps, creating an algorithm leading to the system Hamiltonian and then its Lagrangian. These steps are present, but not always explicit in standard textbooks on electrodynamics. Further, in Sec.~\ref{sec:Space}, we follow exactly the same steps starting with the same Maxwell's equations, but choosing the spatial direction $z$ for  the evolution instead of time. As a result, we arrive at a spatial evolution generator in an unambiguous way and clarify its physical meaning. In a similar way, we construct a spatial Lagrangian of the electromagnetic field with sources. Section~\ref{sec:Conclusion} concludes the paper.

\section{Temporal evolution \label{sec:Time}}

In this section we briefly reproduce the main relations of the conventional approach to the dynamics of the electromagnetic field with sources, where the evolution is considered in time. Our argumentation follows closely that of Cohen-Tannoudji, Dupont-Roc, and Grynberg \cite{Cohen-Tannoudji89}.

We start with Maxwell's equations for the electric field $\mathbf{E}(\mathbf{r},t)$ and the magnetic field $\mathbf{B}(\mathbf{r},t)$, which are real vector functions of the spatial coordinate $\mathbf{r}=(x,y,z)$ and time $t$:
\begin{eqnarray}\label{Maxwell1}
\nabla\,\mathbf{E}(\mathbf{r},t) &=& \frac1{\epsilon_0}\rho(\mathbf{r},t),\\\label{Maxwell2}
\nabla\,\mathbf{B}(\mathbf{r},t) &=& 0,\\\label{Maxwell3}
\nabla \times \mathbf{E}(\mathbf{r},t) &=&  - \frac{\partial}{\partial t}\mathbf{B}(\mathbf{r},t),\\\label{Maxwell4}
\nabla \times \mathbf{B}(\mathbf{r},t) &=&  \frac1{\epsilon_0c^2}\mathbf{J}(\mathbf{r},t) + \frac1{c^2}\frac{\partial}{\partial t}\mathbf{E}(\mathbf{r},t).
\end{eqnarray}
Here $\rho(\mathbf{r},t)$ is the scalar field of charge density, $\mathbf{J}(\mathbf{r},t)$ is the vector field of the current, $c$ is the speed of light in vacuum, and $\epsilon_0$ is the vacuum permittivity. The charge density and the current satisfy the equation of continuity
\begin{equation}\label{Cont}
\frac{\partial}{\partial t}\rho(\mathbf{r},t) + \nabla\,\mathbf{J}(\mathbf{r},t) = 0.
\end{equation}

\subsection{\textbf{k} space}
A three-dimensional Fourier transform of all four fields with respect to the spatial coordinate $\mathbf{r}$ gives us
\begin{eqnarray}\label{barE}
\bar{\mathbf{\mathcal{E}}}(\mathbf{k},t) &=& \frac{1}{(2\pi)^\frac32}\int dx \int dy \int dz \mathbf{E}(\mathbf{r},t) e^{-i\mathbf{kr}},\\\nonumber
\end{eqnarray}
with similar expressions for
$\bar{\mathbf{\mathcal{B}}}(\mathbf{k},t)$, $\bar{\mathcal{\rho}}(\mathbf{k},t)$, and $\bar{\mathbf{\mathcal{J}}}(\mathbf{k},t)$. Here $\mathbf{k}=(k_x,k_y,k_z)$ is the wave vector and all three  integrals are taken from $-\infty$ to $\infty$. All the fields are real in the direct domain, but their Fourier transforms are complex and satisfy
\begin{equation}\label{lindep}
\bar{\mathbf{\mathcal{E}}}(-\mathbf{k},t) = \bar{\mathbf{\mathcal{E}}}^*(\mathbf{k},t),
\end{equation}
and similar relations for the other three fields.

Multiplying Eqs. (\ref{Maxwell1}) - (\ref{Maxwell4}) by $e^{-i\mathbf{kr}}$ and integrating over $\mathbf{r}$ we obtain Maxwell's equations in $\mathbf{k}$ space:
\begin{eqnarray}\label{Maxwell1k}
i\mathbf{k}\bar{\mathbf{\mathcal{E}}}(\mathbf{k},t) &=& \frac1{\epsilon_0}\bar{\rho}(\mathbf{k},t),\\\label{Maxwell2k}
i\mathbf{k}\bar{\mathbf{\mathcal{B}}}(\mathbf{k},t) &=& 0,\\\label{Maxwell3k}
i\mathbf{k} \times \bar{\mathbf{\mathcal{E}}}(\mathbf{k},t) &=&  - \frac{\partial}{\partial t}\bar{\mathbf{\mathcal{B}}}(\mathbf{k},t),\\\label{Maxwell4k}
i\mathbf{k} \times \bar{\mathbf{\mathcal{B}}}(\mathbf{k},t) &=&  \frac1{\epsilon_0c^2}\bar{\mathbf{\mathcal{J}}}(\mathbf{k},t) + \frac1{c^2}\frac{\partial}{\partial t}\bar{\mathbf{\mathcal{E}}}(\mathbf{k},t).
\end{eqnarray}

The equation of continuity, Eq. (\ref{Cont}), takes the form
\begin{equation}\label{Contk}
\frac{\partial}{\partial t}\bar{\rho}(\mathbf{k},t) + i\mathbf{k}\bar{\mathbf{\mathcal{J}}}(\mathbf{k},t) = 0.
\end{equation}

\subsection{Normal variables \label{sec:Time-normal}}

For a given wave vector $\mathbf{k}$ we split the electric field into the longitudinal part $\bar{\mathbf{\mathcal{E}}}_\parallel(\mathbf{k},t)$ (parallel to $\mathbf{k}$) and the transverse part $\bar{\mathbf{\mathcal{E}}}_\perp(\mathbf{k},t)$ (orthogonal to $\mathbf{k}$): $\bar{\mathbf{\mathcal{E}}}(\mathbf{k},t) = \bar{\mathbf{\mathcal{E}}}_\parallel(\mathbf{k},t) + \bar{\mathbf{\mathcal{E}}}_\perp(\mathbf{k},t)$, with similar expressions for the fields $\bar{\mathbf{\mathcal{B}}}(\mathbf{k},t)$ and $\bar{\mathbf{\mathcal{J}}}(\mathbf{k},t)$. Equations (\ref{Maxwell1k}) and (\ref{Maxwell2k}) imply that the magnetic field is purely transverse while the longitudinal part of the electric field is
\begin{equation}\label{Elong}
\bar{\mathbf{\mathcal{E}}}_\parallel(\mathbf{k},t) = -\frac{i\boldsymbol{\kappa}}{\epsilon_0k}\bar{\rho}(\mathbf{k},t),
\end{equation}
where $k=|\mathbf{k}|$ and $\boldsymbol{\kappa}=\mathbf{k}/k$ is the unit vector in the direction of the wave vector $\mathbf{k}$.

Now, we define the three-dimensional vector field
\begin{equation}\label{alpha}
\boldsymbol{\alpha}(\mathbf{k},t) = -i\sqrt{\frac{\epsilon_0}{2\hbar c k}} \left[\bar{\mathbf{\mathcal{E}}}_\perp(\mathbf{k},t) - c\boldsymbol{\kappa} \times \bar{\mathbf{\mathcal{B}}}(\mathbf{k},t)\right].
\end{equation}
Note that $\boldsymbol{\alpha}(\mathbf{k},t)$ is linearly independent of its complex conjugate $\boldsymbol{\alpha}^*(\mathbf{-k},t)$, though the corresponding quantities for the electric and magnetic fields are linearly dependent due to relations similar to Eq. (\ref{lindep}). By definition, $\boldsymbol{\alpha}(\mathbf{k},t)$ is transverse, and therefore can be decomposed as
\begin{equation}\label{alphavector}
\boldsymbol{\alpha}(\mathbf{k},t) = \alpha_1(\mathbf{k},t) \mathbf{n}_1(\mathbf{k}) + \alpha_2(\mathbf{k},t) \mathbf{n}_2(\mathbf{k}),
\end{equation}
where $\mathbf{n}_1(\mathbf{k})$ and $\mathbf{n}_2(\mathbf{k})$ are two unit vectors orthogonal to $\mathbf{k}$, with $\alpha_1(\mathbf{k},t)$ and $\alpha_2(\mathbf{k},t)$ as its two vector components. 

\begin{figure}[h]
\centering
\includegraphics[width=\columnwidth]{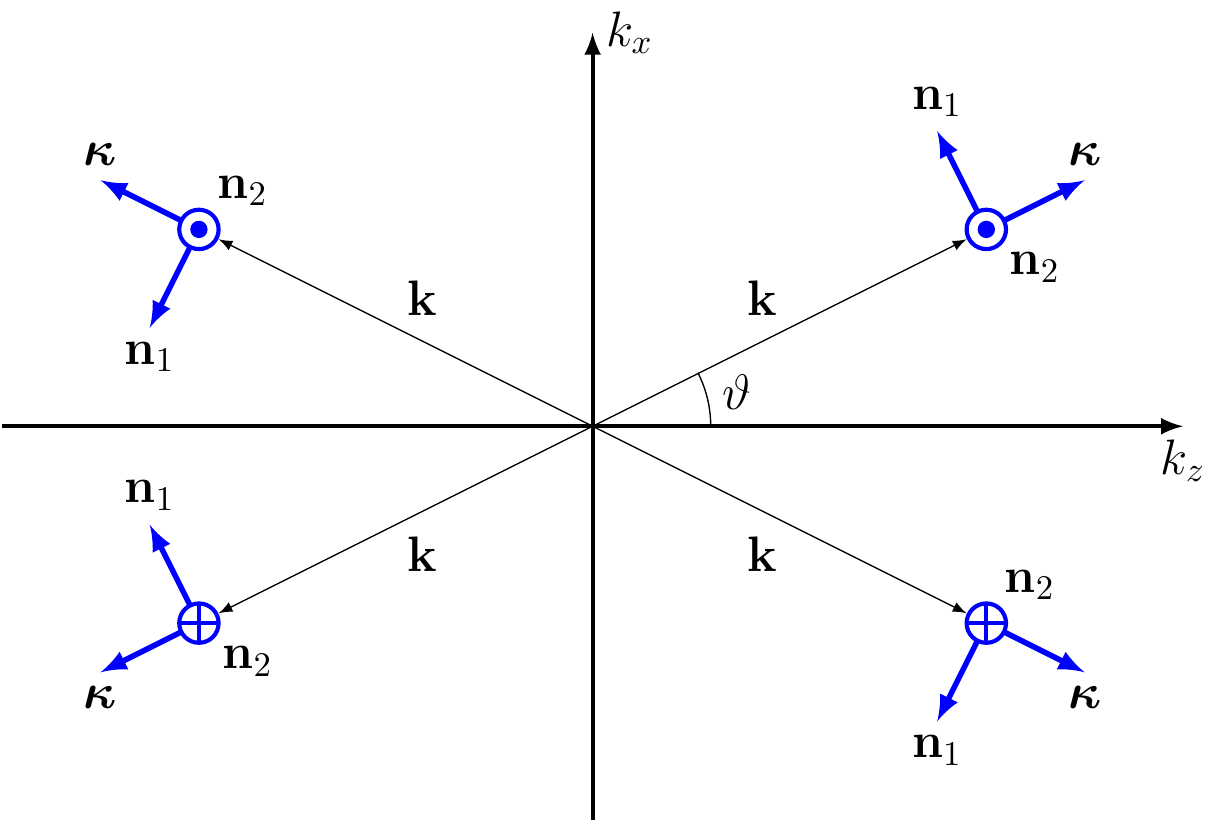}
\caption{Basis vectors for a given wave vector $\mathbf{k}$, lying in the $(k_xk_z)$ plane. The direction of the basis vectors is specific in each of the four quadrants of this plane. \label{fig:Basis}}
\end{figure}

To be specific, we define the vector $\mathbf{n}_1(\mathbf{k})$ so that it lies in the plane, containing $\mathbf{k}$ and the $z$ axis, and its $z$ component is always nonpositive (see Fig.~\ref{fig:Basis}). The other vector is defined as $\mathbf{n}_2(\mathbf{k})=\boldsymbol{\kappa}\times\mathbf{n}_1(\mathbf{k})$.  Introducing spherical coordinates with the polar angle $\vartheta(\mathbf{k})=\cos^{-1}\left(k_z/k\right)$ and the azimuthal angle $\varphi(\mathbf{q})$ such that $\cos\varphi(\mathbf{q})=k_x/q$ and $\sin\varphi(\mathbf{q})=k_y/q$, we write the three basis vectors as
\begin{eqnarray}\label{kappa}
\boldsymbol{\kappa}(\mathbf{k}) &=& \cos\vartheta(\mathbf{k})\mathbf{n}_z + \sin\vartheta(\mathbf{k})\mathbf{n}_q,\\\label{n1}
\mathbf{n}_1(\mathbf{k}) &=& -\sin\vartheta(\mathbf{k})\mathbf{n}_z + \cos\vartheta(\mathbf{k})\mathbf{n}_q,\\\label{n2}
\mathbf{n}_2(\mathbf{k}) &=& \cos\varphi(\mathbf{q})\mathbf{n}_y - \sin\varphi(\mathbf{q})\mathbf{n}_x,
\end{eqnarray}
where $\mathbf{n}_{x,y,z}$ are unit vectors in the directions of the corresponding axes, $\mathbf{q}=(k_x,k_y)$ is the transverse wave vector of length $q=\sqrt{k_x^2+k_y^2}$, and $\mathbf{n}_q=\mathbf{q}/q$ is the unit vector in its direction. The basis vectors defined in this way possess the symmetry properties $\mathbf{n}_1(-\mathbf{k})=\mathbf{n}_1(\mathbf{k})$ and $\mathbf{n}_2(-\mathbf{k})=-\mathbf{n}_2(\mathbf{k})$.

Differentiating both sides of Eq.~(\ref{alpha}) by time and using Eqs. (\ref{Maxwell3k}) and (\ref{Maxwell4k}), we obtain the following equation of motion
\begin{equation}\label{alphaeq}
\frac{\partial}{\partial t}\boldsymbol{\alpha}(\mathbf{k},t) = -ick \boldsymbol{\alpha}(\mathbf{k},t) + \frac{i}{\sqrt{2\epsilon_0\hbar ck}} \bar{\mathbf{\mathcal{J}}}_\perp(\mathbf{k},t),
\end{equation}
which can be considered as a system of two equations for the components $\alpha_1(\mathbf{k},t)$ and $\alpha_2(\mathbf{k},t)$. This system is diagonal for the field variables, in the sense that the derivative of each variable is expressed through the same variable, and not through the other, i.e. the system matrix is diagonal. Such variables are known as ``normal coordinates'' in theoretical mechanics \cite{Goldstein80,Landau-LifshitzI}. Following Ref.~\cite{Cohen-Tannoudji89}, we call $\alpha_1(\mathbf{k},t)$ and $\alpha_2(\mathbf{k},t)$ ``complex normal variables'' of the electromagnetic field, to distinguish them from the generalized coordinates of the Hamiltonian formalism. Physically, they are amplitudes of plane waves, which are spatial normal modes of the free field.

Real normal variables $\bar{X}_j(\mathbf{k},t)$ and $\bar{Y}_j(\mathbf{k},t)$ can be defined by the decomposition 
\begin{equation}\label{Xbar}
\alpha_j(\mathbf{k},t) = \sqrt{\frac{\epsilon_0}{2\hbar ck}}\left(ck\bar{X}_j(\mathbf{k},t)+\frac{i}{\epsilon_0}\bar{Y}_j(\mathbf{k},t)\right).
\end{equation}
Substituting Eq.~(\ref{Xbar}) into Eq.~(\ref{alphaeq}) and equalizing the real and imaginary parts of both sides, we obtain first-order equations for  $\bar{X}_j(\mathbf{k},t)$ and $\bar{Y}_j(\mathbf{k},t)$. Then, differentiating them by time and excluding
$\partial_t\bar{Y}_j(\mathbf{k},t)$, we obtain a system of second-order differential equations for $\bar{X}_j(\mathbf{k},t)$, which is diagonal for these variables. A similar diagonal system can be found for the variables $\bar{Y}_j(\mathbf{k},t)$. For this reason $\bar{X}_j(\mathbf{k},t)$ and $\bar{Y}_j(\mathbf{k},t)$ can be considered as two sets of real normal variables of the field.

\subsection{Expressing the fields via the normal variables}

By inverting Eq.~(\ref{alpha}), we obtain the transverse part of the electric field and the magnetic field expressed via the complex normal variables as 
\begin{equation}\label{Etrans}
\mathbf{E}_\perp(\mathbf{r},t) = ic\int d^3kA_kk\sum_{j=1}^2\left[\alpha_j(\mathbf{k},t)e^{i\mathbf{kr}} \mathbf{n}_j(\mathbf{k})\right. - \left.c.c.\right]
\end{equation}
and
\begin{equation}\label{Btrans}
\mathbf{B}(\mathbf{r},t) = i\int d^3kA_kk\sum_{j=1}^2\left[\alpha_j(\mathbf{k},t)e^{i\mathbf{kr}} \boldsymbol{\kappa}\times\mathbf{n}_j(\mathbf{k})\right. - \left.c.c.\right],
\end{equation}
with
\begin{equation}\label{Ak}
A_k = \sqrt{\frac{\hbar}{2\epsilon_0(2\pi)^3ck}}.
\end{equation}
The longitudinal part of the electric field is given by the inverse Fourier transform of Eq. (\ref{Elong}).

The vector potential $\mathbf{A}(\mathbf{r},t)$ and the scalar potential $\Phi(\mathbf{r},t)$ fields are introduced by the relations
\begin{eqnarray}\label{EvsA}
\mathbf{E}(\mathbf{r},t) &=& -\frac{\partial}{\partial t} \mathbf{A}(\mathbf{r},t) - \nabla \Phi(\mathbf{r},t),\\\label{BvsA}
\mathbf{B}(\mathbf{r},t) &=& \nabla \times \mathbf{A}(\mathbf{r},t).
\end{eqnarray}
In the Coulomb gauge, the vector potential is transverse and is expressed via the normal variables as
\begin{equation}\label{A}
\mathbf{A}(\mathbf{r},t) = \int d^3kA_k\sum_{j=1}^2\left[\alpha_j(\mathbf{k},t)e^{i\mathbf{kr}} \mathbf{n}_j(\mathbf{k}) + c.c.\right].
\end{equation}
This relation is compatible with Eqs.~(\ref{alphaeq}), (\ref{Etrans}), and (\ref{EvsA}) due to the symmetry property of the current field $\bar{\mathbf{\mathcal{J}}}_\perp(-\mathbf{k},t)=\bar{\mathbf{\mathcal{J}}}_\perp(\mathbf{k},t)^*$. 

\subsection{Hamiltonian equation}

The equation for the normal variables, Eq. (\ref{alphaeq}), can be rewritten in the Hamiltonian form
\begin{equation}\label{Hamilteq}
\frac{\partial}{\partial t}\alpha_j(\mathbf{k},t) = \left\{ \alpha_j(\mathbf{k},t),H \right\},
\end{equation}
where $\left\{...\right\}$ stands for the Poisson bracket, which is defined for any two functionals $U$ and $V$ of the normal variables $\alpha_1(\mathbf{k},t)$, $\alpha_2(\mathbf{k},t)$, and their complex conjugates as \cite{Cohen-Tannoudji89}
\begin{eqnarray}\label{Poisson}
\left\{ U,V \right\}  
= &-&\frac{i}\hbar\int\sum_{j=1}^2\left(\frac{\delta U}{\delta \alpha_j(\mathbf{k},t)}\frac{\delta V}{\delta \alpha_j^*(\mathbf{k},t)}\right.\\\nonumber
&-& \left.\frac{\delta U}{\delta \alpha_j^*(\mathbf{k},t)}\frac{\delta V}{\delta \alpha_j(\mathbf{k},t)}\right)d^3k.
\end{eqnarray}
Here, as usual in the complex Hamiltonian formalism, $\alpha_j(\mathbf{k},t)$ and $\alpha_j^*(\mathbf{k},t)$ are formally considered as two independent variables, and $\delta U/\delta\alpha$ means the functional derivative, see details in Appendix~\ref{appendix:Poisson}.

Choosing $U=\alpha_j(\mathbf{k},t)$, $V=\alpha_l^*(\mathbf{k}',t)$ or $V=\alpha_l(\mathbf{k}',t)$, we arrive at the canonical Poisson brackets
\begin{eqnarray}\label{Poisson1}
\left\{\alpha_j(\mathbf{k},t),\alpha_l^*(\mathbf{k}',t)\right\} &=& -\frac{i}\hbar\delta_{jl}\delta(\mathbf{k}-\mathbf{k}'),\\\label{Poisson2}
\left\{\alpha_j(\mathbf{k},t),\alpha_l(\mathbf{k}',t)\right\} &=& 0.
\end{eqnarray}

The Hamiltonian $H$ in Eq. (\ref{Hamilteq}) can be written in two different forms, depending on whether the movement of charges depends on the electromagnetic field or not. 

\subsubsection{Particles' movements depend on the field}
In the first case, it is convenient to consider particles numbered by index $\mu$, each having mass $m_\mu$, charge $q_\mu$, position $\mathbf{r}_\mu(t)$, and velocity $\mathbf{v}_\mu(t)$. The dynamics of these particles is governed by the Newton-Lorentz equations
\begin{equation}\label{NL}
m_\mu\frac{d^2}{dt^2}\mathbf{r}_\mu(t) = q_\mu \mathbf{E}(\mathbf{r}_\mu(t),t) + q_\mu \mathbf{v}_\mu(t) \times \mathbf{B}(\mathbf{r}_\mu(t),t).
\end{equation}
The charge density and the current density are expressed as
\begin{eqnarray}\label{rho}
\rho(\mathbf{r},t) &=& \sum_\mu q_\mu\delta\left[\mathbf{r}-\mathbf{r}_\mu(t)\right],\\\label{j}
\mathbf{J}(\mathbf{r},t) &=& \sum_\mu q_\mu \mathbf{v}_\mu(t) \delta\left[\mathbf{r}-\mathbf{r}_\mu(t)\right],
\end{eqnarray}
and satisfy the continuity equation, Eq. (\ref{Cont}).

The Newton-Lorentz equation, Eq.~(\ref{NL}), can be rewritten in the Hamiltonian form
\begin{equation}\label{NLH}
\frac{d}{dt}\mathbf{r}_\mu(t) = \left\{\mathbf{r}_\mu(t),H\right\},\hspace{5mm}
\frac{d}{dt}\mathbf{p}_\mu(t) = \left\{\mathbf{p}_\mu(t),H\right\},
\end{equation}
where $\mathbf{p}_\mu(t) = m_\mu\mathbf{v}_\mu(t) +q_\mu\mathbf{A}(\mathbf{r}_\mu,t)$ is the generalized momentum of the $\mu$th particle. The Hamiltonian giving both Eqs.~(\ref{NLH}) and (\ref{Hamilteq}) is a sum of four terms,
\begin{equation}\label{H}
H=H_F+H_P+H_\mathrm{int}+H_\mathrm{long},
\end{equation}
where
\begin{equation}\label{HF}
H_F = \int \hbar ck\sum_{j=1}^2\alpha_j^*(\mathbf{k},t)\alpha_j(\mathbf{k},t)d^3k
\end{equation}
is the Hamiltonian of the free transverse field,
\begin{equation}\label{HP}
H_P = \sum_\mu\frac{\mathbf{p}_\mu^2(t)}{2m_\mu}
\end{equation}
is the Hamiltonian of free particles, and
\begin{equation}\label{Hint}
H_\mathrm{int} = \sum_\mu\frac1{2m_\mu}\left[q_\mu^2\mathbf{A}^2(\mathbf{r}_\mu,t)
-2q_\mu\mathbf{p}_\mu(t)\mathbf{A}(\mathbf{r}_\mu,t)\right]
\end{equation}
is the Hamiltonian of their interaction, the argument of $\mathbf{r}_\mu(t)$ being omitted for compactness: $\mathbf{r}_\mu=\mathbf{r}_\mu(t)$. 

The last term in Eq.~(\ref{H}), 
\begin{equation}\label{Hlong}
H_\mathrm{long} = \frac{\epsilon_0}2 \int \left|\bar{\mathbf{\mathcal{E}}}_\parallel(\mathbf{k},t)\right|^2d^3k,
\end{equation}
is the energy of the longitudinal field. The latter is directly related to the charge density via Eq.~(\ref{Elong}) and does not represent a dynamical variable. It can be shown that $H_\mathrm{long}$ is the Coulomb electrostatic energy of the system of charges \cite{Cohen-Tannoudji89}.

Equations~(\ref{alphaeq}) and (\ref{NL}) are obtained as Hamiltonian equations from the Hamiltonian, Eq.~(\ref{H}), by applying the canonical equal-time Poisson brackets for the field, Eqs.~(\ref{Poisson1}) and (\ref{Poisson2}), and similar relations for the particles
\begin{equation}\label{Poisson3}
\left\{r_{\mu j}(t),p_{\nu l}(t)\right\} = \delta_{\mu\nu}\delta_{jl},
\end{equation}
with all other Poisson brackets for the dynamical variables $\left(\alpha_j(\mathbf{k},t),\alpha_l^*(\mathbf{k},t),\mathbf{r}_\mu(t),\mathbf{p}_\nu(t)\right)$ being zero. In Eq.~(\ref{Poisson3}) $r_{\mu j}$ and $p_{\nu l}$ are the $j$th and $l$th components of the vectors $\mathbf{r}_\mu$ and $\mathbf{p}_\nu$ respectively.

\subsubsection{Particles' movements do not depend on the field \label{sec:indep}}
A different situation is met when the distribution and motion of the charges do not depend on the field. They may be predetermined by some external forces \cite{Glauber63b} or the effect of the field on the charges may be negligible \cite{Akhiezer65}. In this case the Hamiltonian is %
\begin{equation}\label{Hprime}
H=H_F+H_\mathrm{int}',
\end{equation}
where $H_F$ is given by Eq.~(\ref{HF}), while the interaction Hamiltonian is
\begin{equation}\label{Hintprime}
H_\mathrm{int}'= -\int \mathbf{J}(\mathbf{r},t)\mathbf{A}(\mathbf{r},t)d^3r.
\end{equation}
Using Eq.~(\ref{A}) we rewrite the interaction Hamiltonian as 
\begin{equation}\label{Hintprime2}
H_\mathrm{int}'= -\int \sqrt{\frac\hbar{2\epsilon_0ck}}\left[ \bar{\mathbf{\mathcal{J}}}_\perp(\mathbf{k},t) \boldsymbol{\alpha}^*(\mathbf{k},t)+c.c.\right]d^3k.
\end{equation}
It is easy to see that Eq.~(\ref{alphaeq}) has the form of Eq.~(\ref{Hamilteq}) with the Hamiltonian $H$ and the Poisson bracket, Eq.~(\ref{Poisson1}).

\subsection{Alternative Hamiltonian \label{sec:AltH}}

The Hamiltonian form of the equation for the normal variables, obtained in the previous section, is not unique. Indeed, it corresponds to identifying, for every $\mathbf{k}$ and $t$, the generalized coordinate with $\bar{X}_j(\mathbf{k},t)$ and the generalized momentum with $\bar{Y}_j(\mathbf{k},t)$. However, an alternative representation exists, where the generalized momentum is identified with $-\bar{Y}_j(\mathbf{k},t)$. This is equivalent to introducing an alternative complex normal variable $\tilde\alpha_j(\mathbf{k},t)=\alpha_j^*(\mathbf{k},t)$, also obeying a diagonal system of first-order differential equations. These equations can be written in the Hamiltonian form 
\begin{equation}\label{Hamilteqalt}
\frac{\partial}{\partial t}\tilde\alpha_j(\mathbf{k},t) = \left\{ \tilde\alpha_j(\mathbf{k},t),\tilde H \right\}_\mathrm{alt},
\end{equation}
where $\tilde H=-H$ and the Poisson bracket $\left\{...\right\}_\mathrm{alt}$ is defined as in Eq.~(\ref{Poisson}) but with respect to $\tilde\alpha_j(\mathbf{k},t)$. 

The choice of generalized momentum is not a pure formality, because, at the quantization stage, it affects the definition of the photon creation and annihilation operators, as we will see below. In standard textbooks on quantum electrodynamics (see, e.g., Refs.~\cite{Cohen-Tannoudji89,Akhiezer65,Mandel-Wolf}), the Hamiltonian $H$ is either deduced from a postulated Lagrangian, or accepted because it corresponds to the system energy, and not to the negative energy, as $\tilde H$, while it is taken for granted that the energy is the generator of the temporal evolution. Neither of these methods is acceptable in our \emph{inductive} treatment, where we do not know in advance what physical quantity plays the role of generator, and we aim at constructing a Lagrangian from the equations of motion. We note that, since a negative momentum corresponds to the time inversion in a mechanical system, the discussed ambiguity in the choice of Hamiltonian is related to the time-inversion invariance of Maxwell's equations.

One consequence of the traditional identification of the energy with the generator of time evolution is the fact, that the positive-frequency part of the electric field is expressed via the photon annihilation operators, and not via the photon creation ones, which is a cornerstone of Glauber's theory of optical coherence \cite{Glauber63a}. Thus, we fix the choice of $H$ as the Hamiltonian and $\alpha_j(\mathbf{k},t)$ as the complex normal variable determining the Poisson bracket by the following postulate. 
\begin{postulate}[on the positive-frequency part]
  The generalized coordinates $q_n$ and momenta $p_n$ in the Hamiltonian form of the equations of motion are chosen in such a way that the positive-frequency parts of the fields are expressed via the combinations $\mu q_n+i\nu p_n$ with positive $\mu$ and $\nu$, and not via their complex conjugates.
\end{postulate}

The complex normal variable $\alpha_j(\mathbf{k},t)$ satisfies this postulate because in the absence of sources it varies with time as $e^{-ickt}$, and therefore belongs to the positive-frequency part of the field. The positive-frequency parts of the fields $\mathbf{E}_\perp(\mathbf{r},t)$, $\mathbf{B}(\mathbf{r},t)$, and $\mathbf{A}(\mathbf{r},t)$  are expressed via $\alpha_j(\mathbf{k},t)$ and not via its complex conjugate, as can be seen from Eqs.~(\ref{Etrans}) and (\ref{A}).

The choice of the generalized coordinates and momenta satisfying the above postulate is still not unique, the complex normal variable can be modified by a phase factor $\alpha_j(\mathbf{k},t)\to\alpha_j(\mathbf{k},t)e^{i\xi(\mathbf{k})}$ with a real $\xi(\mathbf{k})$. However, such a modification corresponds to a canonical transformation and therefore does not change the Poisson bracket. Note, that the transformation $(q,p)\to(q,-p)$ discussed above is not a canonical transformation and inverts the sign of the Poisson bracket.

\subsection{Expressing the Hamiltonian via the fields \label{sec:Time-H}}

Using the definition of complex normal variables, Eq.~(\ref{alpha}), we can rewrite the Hamiltonian of the free transverse field directly through the electric and magnetic fields: 
\begin{equation}\label{HF2}
H_F = \frac{\epsilon_0}{2}\int \left[\mathbf{E}^2_\perp(\mathbf{r},t)+c^2\mathbf{B}^2(\mathbf{r},t)\right]d^3r.
\end{equation}

It can be shown that the total Hamiltonian $H$ represents, up to an additive constant, the full energy of the field and particles. For brevity, we show it only for the case of particles, independent of the field, the other case being treated in Ref.~\cite{Cohen-Tannoudji89}. 

The transverse electric field under the integral in Eq.~(\ref{HF2}) can be replaced by the full electric field, since the longitudinal electric field does not constitute a dynamical variable. Thus, we write
\begin{equation}\label{HF3}
H_F = \int T^{00}(\mathbf{r},t)d^3r,
\end{equation}
where
\begin{equation}\label{T00F}
T^{00}(\mathbf{r},t) = \frac{\epsilon_0}{2} \left[\mathbf{E}^2(\mathbf{r},t)+c^2\mathbf{B}^2(\mathbf{r},t)\right]
\end{equation}
is the energy density of the field, which is an element of the energy-momentum tensor of the field $T^{\mu\nu}$ \cite{Landau-LifshitzII,Jackson}. 

In a similar way we express the interaction Hamiltonian $H_\mathrm{int}'$ as an integral of the interaction energy density 
\begin{equation}\label{T00I}
T^{00}_I(\mathbf{r},t) =  - \mathbf{J}(\mathbf{r},t)\mathbf{A}(\mathbf{r},t).
\end{equation}
The full energy density of field and sources is defined in general as $T^{00}_{FS}=T^{00}+T^{00}_I+T^{00}_S$, where $T^{00}_S$ is the energy density of the sources. In the considered case of particles independent of the field, the latter can be omitted, since it does not contain dynamical variables. Thus, the full Hamiltonian, Eq.~(\ref{Hprime}), can be written as the full energy
\begin{equation}\label{Hfull}
H = \int T^{00}_{FS}(\mathbf{r},t)d^3r.
\end{equation}

\subsection{Lagrangian}

We have seen in the previous sections that Maxwell's equations can be written in a Hamiltonian form. Here we show that they can be recast in a Lagrangian form. For this purpose, for every $\mathbf{k}$, $t$ and $j$, we choose two generalized coordinates
\begin{eqnarray}\label{Qbar}
\bar{Q}_{j1}(\mathbf{k},t)&=&\frac1{\sqrt{2}} \left[\bar{X}_j(\mathbf{k},t)+\bar{X}_j(-\mathbf{k},t)\right],\\\nonumber
\bar{Q}_{j2}(\mathbf{k},t)&=&\frac1{\sqrt{2}ck\epsilon_0} \left[\bar{Y}_j(\mathbf{k},t)-\bar{Y}_j(-\mathbf{k},t)\right],
\end{eqnarray}
where $\bar X_j$ and $\bar Y_j$ are defined by Eq.~(\ref{Xbar}). These coordinates are linearly independent in the half-space $k_z>0$ and are obtained from the coordinates of the Hamiltonian formalism by a canonical transformation, aiming at removing the sources from the first-order equations for the new generalized coordinates due to the condition $\bar{\mathbf{\mathcal{J}}}_\perp(-\mathbf{k},t)=\bar{\mathbf{\mathcal{J}}}_\perp(\mathbf{k},t)^*$. The corresponding momenta are 
\begin{eqnarray}\label{Pbar}
\bar{P}_{j1}(\mathbf{k},t)&=&\frac1{\sqrt{2}} \left[\bar{Y}_j(\mathbf{k},t)+\bar{Y}_j(-\mathbf{k},t)\right],\\\nonumber
\bar{P}_{j2}(\mathbf{k},t)&=&-\frac{ck\epsilon_0}{\sqrt{2}} \left[\bar{X}_j(\mathbf{k},t)-\bar{X}_j(-\mathbf{k},t)\right],
\end{eqnarray}
with the standard Poisson bracket $\left\{\bar{Q}_{jn}(\mathbf{k},t),\bar{P}_{lm}(\mathbf{k}',t)\right\}=\delta_{jl}\delta_{nm}\delta(\mathbf{k}-\mathbf{k}')$. The equations of motion, Eq.~(\ref{alphaeq}), in the new coordinates take the form
\begin{eqnarray}\label{eqQPbar}
\frac{\partial}{\partial t}\bar{Q}_{jn}(\mathbf{k},t) &=& \frac1{\epsilon_0} \bar{P}_{jn}(\mathbf{k},t),\\\nonumber
\frac{\partial}{\partial t}\bar{P}_{j1}(\mathbf{k},t) &=& -c^2k^2\epsilon_0 \bar{Q}_{j1}(\mathbf{k},t)+\sqrt{2}\Rea\bar{\mathcal{J}}_{\perp j}(\mathbf{k},t),\\\nonumber
\frac{\partial}{\partial t}\bar{P}_{j2}(\mathbf{k},t) &=& -c^2k^2\epsilon_0 \bar{Q}_{j2}(\mathbf{k},t)+\sqrt{2}\Ima\bar{\mathcal{J}}_{\perp j}(\mathbf{k},t),
\end{eqnarray}
coinciding with that of a set of forced harmonic oscillators \cite{Goldstein80,Landau-LifshitzI}. In the new variables, the field  Hamiltonian is
\begin{equation}\label{Hnew}
H_{F} =  \fint \sum_{j,n=1}^2\left[\frac1{2\epsilon_0} \bar{P}_{jn}(\mathbf{k},t)^2+\frac{c^2k^2\epsilon_0}2 \bar{Q}_{jn}(\mathbf{k},t)^2\right]d^3k,
\end{equation}
where the integration is taken over half-space $k_z>0$. The interaction Hamiltonian, Eq.~(\ref{Hintprime}), in the new variables is 
\begin{eqnarray}\nonumber
H_\mathrm{int}' =  &-&\frac1{\sqrt{2}}\fint \sum_{j=1}^2 \bar{\mathcal{J}}_{\perp j}(\mathbf{k},t) \left[\bar{Q}_{j1}(\mathbf{k},t)-i\bar{Q}_{j2}(\mathbf{k},t)\right]d^3k \\\label{Hintnew}
&+& c.c.
\end{eqnarray}

The above formalism can be significantly simplified by introducing complex generalized coordinates and momenta \cite{Cohen-Tannoudji89} (not to be confused with the complex normal variables)
\begin{eqnarray}\label{Acoord}
\bar{\mathcal{A}}_j(\mathbf{k},t) &=& \frac{\bar{Q}_{j1}(\mathbf{k},t)+i\bar{Q}_{j2}(\mathbf{k},t)}{\sqrt{2}},\\\label{pibar}
\bar{\pi}_j(\mathbf{k},t) &=& \frac{\bar{P}_{j1}(\mathbf{k},t)+i\bar{P}_{j2}(\mathbf{k},t)}{\sqrt{2}}.
\end{eqnarray}

It can be easily verified, that the complex generalized coordinate coincides with the vector potential in $\mathbf{k}$ space, and that the following Poisson brackets hold: $\left\{\bar{\mathcal{A}}_j(\mathbf{k},t),\bar{\pi}^*_l(\mathbf{k}',t)\right\} =\delta_{jl} \delta\left(\mathbf{k}-\mathbf{k}'\right)$ and $\left\{\bar{\mathcal{A}}_j(\mathbf{k},t),\bar{\pi}_l(\mathbf{k}',t)\right\} = 0$, where the values of $\mathbf{k}$ are restricted to half-space $k_z>0$. 

The Hamiltonian, Eq.~(\ref{Hprime}), has in the new variables the form
\begin{equation}\label{Hprime2}
H = \fint \mathcal{H}(\mathbf{k},t)d^3k,
\end{equation}
where the Hamiltonian density is 
\begin{eqnarray}\nonumber
\mathcal{H}(\mathbf{k},t) &=& \frac1{\epsilon_0} \sum_j\bar{\pi}_j(\mathbf{k},t)\bar{\pi}_j^*(\mathbf{k},t)\\\label{Hdensity}
&+&\epsilon_0c^2k^2\sum_j\bar{\mathcal{A}}_j(\mathbf{k},t)\bar{\mathcal{A}}_j^*(\mathbf{k},t)\\\nonumber
&-& \sum_j \left[\bar{\mathcal{J}}_j^*(\mathbf{k},t)\bar{\mathcal{A}}_j(\mathbf{k},t) + \bar{\mathcal{J}}_j(\mathbf{k},t)\bar{\mathcal{A}}_j^*(\mathbf{k},t)\right].
\end{eqnarray}

The Lagrangian is defined in a standard way 
\begin{equation}\label{Lprime}
L = \fint \mathcal{L}(\mathbf{k},t)d^3k
\end{equation}
with the Lagrangian density \cite{Cohen-Tannoudji89}
\begin{eqnarray}\nonumber
\mathcal{L}(\mathbf{k},t) &=&  \sum_j\left[\bar{\pi}_j(\mathbf{k},t)\dot{\bar{\mathcal{A}}}^*_j(\mathbf{k},t) +\bar{\pi}^*_j(\mathbf{k},t)\dot{\bar{\mathcal{A}}}_j(\mathbf{k},t)\right]\\\label{LdensityDef}
&-& \mathcal{H}(\mathbf{k},t),
\end{eqnarray}
where the conjugate momentum should be expressed via the time derivative of the generalized coordinate. Substituting the Hamiltonian density from Eq.~(\ref{Hdensity}), we find

\begin{eqnarray}\nonumber
\mathcal{L}(\mathbf{k},t) &=& \epsilon_0 \sum_j\dot{\bar{\mathcal{A}}}_j(\mathbf{k},t)\dot{\bar{\mathcal{A}}}^*_j(\mathbf{k},t)\\\label{Ldensity}
&-&\epsilon_0c^2k^2\sum_j\bar{\mathcal{A}}_j(\mathbf{k},t)\bar{\mathcal{A}}_j^*(\mathbf{k},t)\\\nonumber
&+& \sum_j \left[\bar{\mathcal{J}}_j^*(\mathbf{k},t)\bar{\mathcal{A}}_j(\mathbf{k},t) + \bar{\mathcal{J}}_j(\mathbf{k},t)\bar{\mathcal{A}}_j^*(\mathbf{k},t)\right],
\end{eqnarray}
where we have used the relation $\bar{\pi}_j(\mathbf{k},t)=\epsilon_0 \dot{\bar{\mathcal{A}}}_j(\mathbf{k},t)$, obtained from Eqs.~(\ref{pibar}), (\ref{alpha}) and (\ref{EvsA}). The same expression for $\bar{\pi}_j(\mathbf{k},t)$ is given by a functional derivative of $L$ with respect to $\dot{\bar{\mathcal{A}}}^*_j(\mathbf{k},t)$, as required for the conjugate momentum.

Passing to the direct space we obtain
\begin{eqnarray}\label{Lprimedir}
L &=& \int \left(\frac{\epsilon_0}2\left[\left(\dot{\mathbf{A}}+\nabla \Phi\right)^2 -c^2\left(\nabla\times\mathbf{A}\right)^2\right]\right.\\\nonumber
&+&\left.\mathbf{J}\mathbf{A}-\rho\Phi\right)d^3r,
\end{eqnarray}
where we have omitted the arguments $(\mathbf{r},t)$ of the fields for compactness. We have also added the longitudinal field and the term $-\rho\Phi$, not containing dynamical variables and not affecting the equations of motion, but appearing if we keep the term $H_\mathrm{long}$ in the Hamiltonian.

The Lagrangian, Eq.~(\ref{Lprimedir}), is known as the standard Lagrangian of the electromagnetic field \cite{Cohen-Tannoudji89}. The Lagrange equations obtained from this Lagrangian give the wave equations for the field, equivalent to Maxwell's equations. At a higher level of generality, a least action principle can be formulated by defining the action as an integral of $L$ over a fixed interval of time. 

\subsection{Field quantization \label{sec:Time-Quant}}

Canonical quantization of a system having a classical analog is performed in a standard way \cite{Dirac58,Dirac64}: the generalized coordinates and momenta become operators with a commutator equal to the Poisson bracket multiplied by $i\hbar$. In this way, the complex normal variables $\alpha_j(\mathbf{k},t)$ and $\alpha_j^*(\mathbf{k},t)$ become operators $a_j(\mathbf{k},t)$ and $a_j^\dagger(\mathbf{k},t)$ with the canonical bosonic equal-time commutation relations
\begin{eqnarray}\label{comm1}
\left[a_j(\mathbf{k},t),a_l^\dagger(\mathbf{k}',t)\right] &=& \delta_{jl}\delta(\mathbf{k}-\mathbf{k}'),\\\label{comm2}
\left[a_j(\mathbf{k},t),a_l(\mathbf{k}',t)\right] &=& 0.
\end{eqnarray}
It follows from the commutation relations that $a_j(\mathbf{k},t)$ and $a_j^\dagger(\mathbf{k},t)$ have the meanings of the photon annihilation and creation operators respectively \cite{Dirac58}. 

The fields and the Hamiltonian also become operators, retaining their expressions through the normal variables.  The Hamilton equation~(\ref{Hamilteq}) is replaced by the Heisenberg equation
\begin{equation}\label{Heisenberg}
i\hbar\frac{\partial}{\partial t}a_j(\mathbf{k},t) = \left[ a_j(\mathbf{k},t),H \right].
\end{equation}
We note, that the order of the operators $a_j$ and $a_j^\dagger$ in a bilinear bosonic Hamiltonian is irrelevant, since changing this order results in an additive constant in the Hamiltonian, not affecting the equation of motion.

The alternative Poisson bracket, discussed in Sec.~\ref{sec:AltH}, leads to a different quantization scheme, where the operators $a_j(\mathbf{k},t)$ and $a_j^\dagger(\mathbf{k},t)$ exchange their roles. Thus, we see that the existence of a classical Hamiltonian formulation is insufficient for a proper quantization, because there are various Hamiltonian formulations, and some of them are not connected by a canonical transformation. An additional rule is necessary, and this rule consists either in identifying the Hamiltonian with the energy, or in accepting the Postulate on the positive-frequency part, formulated in Sec.~\ref{sec:AltH}. These considerations may seem trivial for the temporal evolution, having just one direction, but become highly nontrivial for the spatial evolution, having two directions along a given spatial axis.

\section{Spatial evolution \label{sec:Space}}

In the previous section, we have shown how the well-known relation between Maxwell's equations and the Hamiltonian or Lagrangian evolution of the field in time can be formulated in an inductive way.  In this section we undertake the same steps starting from the same equations, but choosing the spatial coordinate $z$ as the dimension of the field evolution and  performing Fourier transforms in the other two spatial coordinates and time. In a description of an experiment, it is natural to choose the $z$ axis in the direction of propagation of an optical beam.

\subsection{$\omega\mathbf{q}$ space \label{sec:Space-Maxwell}}

We start by defining a three-dimensional Fourier transform of the electric field
\begin{eqnarray}\label{Eq}
\mathbf{\mathcal{E}}(\omega,\mathbf{q},z) &=& \frac{1}{(2\pi)^\frac32}\int dx \int dy \int dt \mathbf{E}(\mathbf{r},t) e^{-i\mathbf{qx} +i\omega t}\\\nonumber
\end{eqnarray}
with similar expressions for
$\mathbf{\mathcal{B}}(\omega,\mathbf{q},z)$, $\mathcal{R}(\omega,\mathbf{q},z)$,  $\mathbf{\mathcal{J}}(\omega,\mathbf{q},z)$, $\mathbf{\mathcal{A}}(\omega,\mathbf{q},z)$ and $\mathcal{F}(\omega,\mathbf{q},z)$, which are the Fourier transforms of 
$\mathbf{B}(\mathbf{r},t)$,
$\rho(\mathbf{r},t)$,
$\mathbf{J}(\mathbf{r},t)$,
$\mathbf{A}(\mathbf{r},t)$, and 
$\Phi(\mathbf{r},t)$ respectively.
Here $\mathbf{x}=(x,y)$ and all three integrals are taken from $-\infty$ to $\infty$. These Fourier transforms satisfy the relation
\begin{equation}\label{lindep2}
\mathbf{\mathcal{E}}(-\omega,-\mathbf{q},z) = \mathbf{\mathcal{E}}^*(\omega,\mathbf{q},z)
\end{equation}
and similar relations for the other fields.

Multiplying Eqs. (\ref{Maxwell1}) - (\ref{Maxwell4}) by $e^{-i\mathbf{qx}+i\omega t}$ and integrating over $\mathbf{x}$ and $t$ we obtain the Maxwell equations in $\omega\mathbf{q}$ space in a form of six differential equations:
\begin{eqnarray}\label{Maxwell1q}
\frac{\partial \mathcal{E}_x(\omega,\mathbf{q},z)}{\partial z} &=& ik_x\mathcal{E}_z(\omega,\mathbf{q},z)+i\omega \mathcal{B}_y(\omega,\mathbf{q},z),\\\label{Maxwell2q}
\frac{\partial \mathcal{E}_y(\omega,\mathbf{q},z)}{\partial z} &=& ik_y\mathcal{E}_z(\omega,\mathbf{q},z)-i\omega \mathcal{B}_x(\omega,\mathbf{q},z),\\\label{Maxwell3q}
\frac{\partial \mathcal{E}_z(\omega,\mathbf{q},z)}{\partial z} &=& -ik_x\mathcal{E}_x(\omega,\mathbf{q},z)-ik_y\mathcal{E}_y(\omega,\mathbf{q},z)\\\nonumber
&+&\frac1{\epsilon_0} \mathcal{R}(\omega,\mathbf{q},z),\\\label{Maxwell4q}
\frac{\partial \mathcal{B}_x(\omega,\mathbf{q},z)}{\partial z} &=& ik_x\mathcal{B}_z(\omega,\mathbf{q},z)-\frac{i\omega}{c^2}\mathcal{E}_y(\omega,\mathbf{q},z)\\\nonumber
&+&\frac1{\epsilon_0 c^2} \mathcal{J}_y(\omega,\mathbf{q},z),\\\label{Maxwell5q}
\frac{\partial \mathcal{B}_y(\omega,\mathbf{q},z)}{\partial z} &=& ik_y\mathcal{B}_z(\omega,\mathbf{q},z)+\frac{i\omega}{c^2}\mathcal{E}_x(\omega,\mathbf{q},z)\\\nonumber
&-&\frac1{\epsilon_0 c^2} \mathcal{J}_x(\omega,\mathbf{q},z),\\\label{Maxwell6q}
\frac{\partial \mathcal{B}_z(\omega,\mathbf{q},z)}{\partial z} &=& -ik_x\mathcal{B}_x(\omega,\mathbf{q},z)-ik_y\mathcal{B}_y(\omega,\mathbf{q},z),
\end{eqnarray}
and two algebraic equations 
\begin{eqnarray}\label{Maxwell7q}
ik_x\mathcal{E}_y(\omega,\mathbf{q},z) -ik_y\mathcal{E}_x(\omega,\mathbf{q},z) &=& i\omega\mathcal{B}_z(\omega,\mathbf{q},z),\\\label{Maxwell8q}
ik_x\mathcal{B}_y(\omega,\mathbf{q},z) -ik_y\mathcal{B}_x(\omega,\mathbf{q},z) &=& -\frac{i\omega}{c^2}\mathcal{E}_z(\omega,\mathbf{q},z)\\\nonumber
&+& \frac1{\epsilon_0c^2}\mathcal{J}_z(\omega,\mathbf{q},z).
\end{eqnarray}

These equations are \emph{local} in $\omega\mathbf{q}$ space: the derivative of a field at point $(\omega,\mathbf{q},z)$ is expressed via this and other fields at the same point. For this reason, below we omit the argument of a field in $\omega\mathbf{q}$ space to simplify the formulas. 

The equation of continuity, Eq. (\ref{Cont}), takes the form
\begin{equation}\label{Contq}
-i\omega\mathbf{\mathcal{R}} +ik_x\mathcal{J}_x +ik_y\mathcal{J}_y 
+\frac{\partial \mathcal{J}_z}{\partial z} = 0.
\end{equation}

Using the algebraic equations, we can exclude two fields from the differential equations. From the symmetry consideration, we choose the fields $\mathcal{E}_z$ and $\mathcal{B}_z$ for the exclusion. The resulting four equations can be written in a matrix form
\begin{equation}\label{matrix}
\frac{\partial}{\partial z} \left[\begin{array}{c}
  \mathcal{E}_x \\\mathcal{E}_y\\c\mathcal{B}_x\\c\mathcal{B}_y
\end{array}\right] = i\mathbb{M} \left[\begin{array}{c}
  \mathcal{E}_x \\\mathcal{E}_y\\c\mathcal{B}_x\\c\mathcal{B}_y
\end{array}\right] +\frac1{\epsilon_0c\omega} \left[\begin{array}{c}
  ck_x\mathcal{J}_z \\ck_y\mathcal{J}_z\\\omega\mathcal{J}_y\\-\omega\mathcal{J}_x
\end{array}\right],
\end{equation}
where the matrix $\mathbb{M}$ is 
\begin{equation}
\mathbb{M} = \frac{c}{\omega}\left[\begin{array}{cccc}
   0  &  0 & k_xk_y & \frac{\omega^2}{c^2}-k_x^2\\
   0  &  0 & k_y^2-\frac{\omega^2}{c^2} & -k_xk_y\\
   -k_xk_y & k_x^2-\frac{\omega^2}{c^2} & 0 & 0 \\
   \frac{\omega^2}{c^2}-k_y^2 & k_xk_y & 0 & 0
\end{array}\right]
\end{equation}
and we notice that it is block-anti-symmetric when split into $2\times2$ blocks. 

\subsection{Normal variables}

Matrix $\mathbb{M}$ has two eigenvalues $\pm K_z$, where 
\begin{equation}
K_z=\sqrt{\omega^2/c^2-k_x^2-k_y^2},
\end{equation}
and each eigenvalue has multiplicity 2. We use a capital letter here to distinguish $K_z$, which is a function of $\omega$ and $\mathbf{q}$, from the $z$ component of the wave vector $k_z$, appearing in Sec.~\ref{sec:Time}. 

With the help of \emph{Mathematica} 10, we find the linear transformation, bringing $\mathbb{M}$ to a diagonal form. In this way we introduce normal variables
\begin{eqnarray}\label{beta1}
\beta_1^{[\pm]} &=& \mp \frac{i\omega}{c\mathcal{N}} \left(k_x\mathcal{E}_x+k_y\mathcal{E}_y\right) -\frac{icK_z}{\mathcal{N}} \left(k_x\mathcal{B}_y-k_y\mathcal{B}_x\right),\\\label{beta2}
\beta_2^{[\pm]} &=& \pm \frac{i\omega}{\mathcal{N}} \left(k_x\mathcal{B}_x+k_y\mathcal{B}_y\right) -\frac{iK_z}{\mathcal{N}} \left(k_x\mathcal{E}_y-k_y\mathcal{E}_x\right),
\end{eqnarray}
where $\mathcal{N}=q\sqrt{2\hbar K_z\omega^2/\epsilon_0c^2}$. Differentiating the normal variables by $z$ and employing Eq.~(\ref{matrix}), we obtain the equations of motion
\begin{eqnarray}\label{beta1diff}
\frac{\partial\beta_1^{[\pm]}}{\partial z} &=& \pm iK_z\beta_1^{[\pm]} +\frac{iK_z}{\epsilon_0c\mathcal{N}}\left(k_x\mathcal{J}_x+k_y\mathcal{J}_y \mp \frac{q^2}{K_z} \mathcal{J}_z \right) ,\\\label{beta2diff}
\frac{\partial\beta_2^{[\pm]}}{\partial z} &=& \pm iK_z\beta_2^{[\pm]} \pm \frac{i\omega}{\epsilon_0c^2\mathcal{N}}\left( k_x\mathcal{J}_y-k_y\mathcal{J}_x\right),
\end{eqnarray}
diagonal in the normal variables, which confirms the correctness of the chosen linear transformation. Note, that the symbolic calculations are verified by the analytical ones and in no way limit the generality of the obtained results.  

In the absence of sources, Eqs.~(\ref{beta1diff}) and (\ref{beta2diff}) have simple solutions
\begin{equation}\label{betafree}
  \beta_{1,2}^{[\pm]}(\omega,\mathbf{q},z) = \beta_{1,2}^{[\pm]}(\omega,\mathbf{q},z_0) e^{\pm iK_z (z-z_0)},
\end{equation}
which allows us to associate $\beta_{1}^{[+]}$ and $\beta_{2}^{[+]}$ with the waves propagating in the positive direction of the $z$ axis, and $\beta_{1}^{[-]}$ and $\beta_{2}^{[-]}$ with those propagating in the negative direction. 

Because of degeneracy of the eigenvalues, any linear combination of $\beta_{1}^{[+]}$ and $\beta_{2}^{[+]}$ or $\beta_{1}^{[-]}$ and $\beta_{2}^{[-]}$ will also be a normal variable. We have chosen the variables, Eqs.~(\ref{beta1}) and (\ref{beta2}), among many other possibilities, as well as their normalization, in order to obtain simple forms for the electric and magnetic fields expressed through them, as shown in the next section.

The normal variables defined in this way satisfy the relation:
\begin{equation}
  \beta_{1,2}^{[\pm]}(-\omega,-\mathbf{q},z) = \beta_{1,2}^{[\mp]*}(\omega,\mathbf{q},z),
\end{equation}
which means that only the variables with $\omega\ge0$ are linearly independent. Moreover, because of locality of the equations of motion, if the sources with $|\omega|<cq$ are zero everywhere and the variables with $|\omega|<cq$ were zero at some point $z_0$,  then such variables remain zero at all other points. In this paper we consider only the sources for which, at any frequency $\omega$, the transverse wave vector $q$ is upper limited by $|\omega|/c$. As a result, we do not consider the variables with $|\omega|<cq$ as the dynamical variables of the field, putting them to zero identically. Thus, $K_z$ is always real and non-negative.

To summarize, the dynamical variables of the field are $\beta_{1,2}^{[\pm]}(\omega,\mathbf{q},z)$ and $\beta_{1,2}^{[\pm]*}(\omega,\mathbf{q},z)$, where $k_x,k_y\in(-\infty,+\infty)$, $\omega\in[cq,+\infty)$.

\subsection{Expressing the fields via the normal variables \label{sec:Space-fields}}

Now we express the components of the electric field through the normal variables with the help of Eqs.~(\ref{Maxwell8q}), (\ref{beta1}) and (\ref{beta2}) and then make the inverse Fourier transform. As a result, we obtain the electric field as a sum of six terms
\begin{eqnarray}\label{Ert}
\mathbf{E}(\mathbf{r},t) &=& \mathbf{E}^{[+]}(\mathbf{r},t) + \mathbf{E}^{[-]}(\mathbf{r},t) + \mathbf{E}^{[s]}(\mathbf{r},t) \\\nonumber &+& \mathbf{E}^{[+]*}(\mathbf{r},t) +
\mathbf{E}^{[-]*}(\mathbf{r},t) + \mathbf{E}^{[s]*}(\mathbf{r},t)
\end{eqnarray}
where $\mathbf{E}^{[+]}(\mathbf{r},t)$ depends on $\beta_{1,2}^{[+]}$ and corresponds to the field propagating in the positive direction of the $z$ axis (right-propagating part), while $\mathbf{E}^{[-]}(\mathbf{r},t)$ depends on $\beta_{1,2}^{[-]}$ and corresponds to the field propagating in the negative direction (left-propagating part). This distinction, denoted by a superscript in square brackets, $[\pm]$, should not be confused with the separation of the positive and negative-frequency parts of the field, traditionally denoted by a superscript in parentheses, $(\pm)$ \cite{Glauber63a,Mandel-Wolf}. Both fields $\mathbf{E}^{[+]}(\mathbf{r},t)$ and $\mathbf{E}^{[-]}(\mathbf{r},t)$ belong to the positive-frequency part. Their complex conjugates $\mathbf{E}^{[+]*}(\mathbf{r},t)$ and $\mathbf{E}^{[-]*}(\mathbf{r},t)$ belong to the negative-frequency part.

The positive-frequency parts are
\begin{eqnarray}\label{Epm}
\mathbf{E}^{[\pm]}(\mathbf{r},t) &=& \frac{i}c\int d\mathbf{q}\int_{cq}^\infty d\omega \sqrt{\frac{\hbar\omega^2}{(2\pi)^3 2\epsilon_0 K_z}} \\\nonumber 
&\times& \left[\beta_{1}^{[\pm]}\mathbf{m}_1(\omega,\pm\mathbf{q}) +\beta_{2}^{[\pm]}\mathbf{m}_2(\mathbf{q})\right] e^{i\mathbf{qx}-i\omega t},
\end{eqnarray}
where the unit vectors $\mathbf{m}_1(\omega,\mathbf{q})$ and $\mathbf{m}_2(\mathbf{q})$ are defined in a manner similar to the vectors $\mathbf{n}_1(\mathbf{k})$ and $\mathbf{n}_2(\mathbf{k})$ of Sec.~\ref{sec:Time-normal}. At each point of $\omega\mathbf{q}$ space satisfying $|\omega|\ge cq$, we introduce spherical coordinates with the polar angle $\theta(\omega,\mathbf{q})=\cos^{-1}\left(cK_z/\omega\right)$ and the azimuthal angle $\varphi(\mathbf{q})$ such that $\cos\varphi(\mathbf{q})=k_x/q$ and $\sin\varphi(\mathbf{q})=k_y/q$, and define a basis of three unit vectors
\begin{eqnarray}\label{varkappa}
\boldsymbol{\varkappa}(\omega,\mathbf{q}) &=& \cos\theta(\omega,\mathbf{q})\mathbf{n}_z + \sin\theta(\omega,\mathbf{q})\mathbf{n}_q,\\\label{m1}
\mathbf{m}_1(\omega,\mathbf{q}) &=& -\sin\theta(\omega,\mathbf{q})\mathbf{n}_z + \cos\theta(\omega,\mathbf{q})\mathbf{n}_q,\\\label{m2}
\mathbf{m}_2(\mathbf{q}) &=& \cos\varphi(\mathbf{q})\mathbf{n}_y - \sin\varphi(\mathbf{q})\mathbf{n}_x.
\end{eqnarray}
Note that the vectors, Eqs.~(\ref{varkappa}), (\ref{m1}), and (\ref{m2}), correspond to the vectors $\boldsymbol{\kappa}(\mathbf{k})$, $\mathbf{n}_1(\mathbf{k})$, and $\mathbf{n}_2(\mathbf{k})$ respectively upon a substitution $\mathbf{k}=(k_x,k_y,\omega K_z/|\omega|)$. For positive frequencies $\omega$, which are exclusively employed for the dynamical variables, $\mathbf{m}_1(\omega,\mathbf{q})$ corresponds to $\mathbf{n}_1(k_x,k_y,K_z)$ belonging to the right-propagating part, while $\mathbf{m}_1(\omega,-\mathbf{q})$ corresponds to $\mathbf{n}_1(-k_x,-k_y,K_z)=\mathbf{n}_1(k_x,k_y,-K_z)$, belonging to the left-propagating one.

The two remaining terms in Eq.~(\ref{Ert}) are 
\begin{eqnarray}\label{Es}
\mathbf{E}^{[s]}(\mathbf{r},t) &=& -i\int d\mathbf{q}\int_{cq}^\infty d\omega \frac{\mathcal{J}_z(\omega,\mathbf{q},z)}{(2\pi)^{3/2}\epsilon_0\omega}\mathbf{n}_z e^{i\mathbf{qx}-i\omega t}
\end{eqnarray}
and its complex conjugate. They show a direct contribution of the current to the $z$ component of the electric field. 

In a similar way, we obtain from Eqs.~(\ref{Maxwell7q}), (\ref{beta1}) and (\ref{beta2}) the magnetic field as a sum of four terms
\begin{eqnarray}\label{Brt}
\mathbf{B}(\mathbf{r},t) &=& \mathbf{B}^{[+]}(\mathbf{r},t) + \mathbf{B}^{[-]}(\mathbf{r},t)\\\nonumber &+& \mathbf{B}^{[+]*}(\mathbf{r},t) +
\mathbf{B}^{[-]*}(\mathbf{r},t),
\end{eqnarray}
where
\begin{eqnarray}\label{Bpm}
\mathbf{B}^{[\pm]}(\mathbf{r},t) &=& \frac{i}{c^2}\int d\mathbf{q}\int_{cq}^\infty d\omega \sqrt{\frac{\hbar\omega^2}{(2\pi)^3 2\epsilon_0 K_z}} \\\nonumber 
&\times& \left[\beta_{1}^{[\pm]}\mathbf{m}_2(\mathbf{q}) -\beta_{2}^{[\pm]}\mathbf{m}_1(\omega,\pm\mathbf{q})\right] e^{i\mathbf{qx}-i\omega t}.
\end{eqnarray}

In the absence of sources we can substitute the solution, Eq.~(\ref{betafree}), into Eq.~(\ref{Epm}) and replace the integral over $\omega$ by one over $K_z$, using the relation $d\omega=c^2K_zdK_z/|\omega|$. In this way, we represent the electric field, Eq.~(\ref{Ert}), as an integral over plain monochromatic waves in $\mathbf{k}$ space. Alternatively, we can solve Eq.~(\ref{alphaeq}) for some initial value $\boldsymbol{\alpha}(\mathbf{k},t_0)$ and substitute the solution into Eq.~(\ref{Etrans}), obtaining another decomposition of the electric field into plain monochromatic waves. The same procedures can be done for the magnetic field. Since the simultaneous decomposition of the electric and magnetic fields into plain monochromatic waves is unique, the coefficients of the decompositions should coincide, which gives us a relation  
\begin{equation}
 \beta_j^{[\pm]}(\omega,\mathbf{q},z_0) = \sqrt{\frac{\omega}{c^2K_z}}\alpha_j(k_x,k_y,\pm K_z,t_0) e^{\pm iK_zz_0+i\omega t_0}. 
\end{equation}

From Eq.~(\ref{Poisson1}) and the standard rule for changing the argument of the delta function, we obtain 
\begin{eqnarray}\nonumber
 \left\{\beta_j^{[\pm]}(\omega,\mathbf{q},z),\beta_l^{[\pm]*}(\omega',\mathbf{q}',z)\right\} &=& -\frac{i}\hbar\delta_{jl}\delta(\omega-\omega')\delta(\mathbf{q}-\mathbf{q}'),\\\label{Poissonbeta}
 \left\{\beta_j^{[\pm]}(\omega,\mathbf{q},z),\beta_l^{[\mp]*}(\omega',\mathbf{q}',z)\right\} &=& 0.
\end{eqnarray}

Thus, in the absence of sources, the transition from the complex normal variables $\alpha_j(\mathbf{k},t)$ to $\beta_j^{[\pm]}(\omega,\mathbf{q},z)$ can be considered as a canonical transformation.

\subsection{Hamiltonian equation}

To write the equations of motion in a Hamiltonian form we define the spatial Poisson bracket in a way similar to Eq.~(\ref{Poisson}): 
\begin{eqnarray}\label{PoissonPrime}
\left\{ U,V \right\}'  
= &-&\frac{i}\hbar\int d\mathbf{q}\int_{cq}^\infty d\omega\\\nonumber &\times&\sum_{j\pm}\left(\frac{\delta U}{\delta \beta_j^{[\pm]}(\omega,\mathbf{q},z)} 
\frac{\delta V}{\delta \beta_j^{[\pm]*}(\omega,\mathbf{q},z)}\right.\\\nonumber
&-& \left.\frac{\delta U}{\delta \beta_j^{[\pm]*}(\omega,\mathbf{q},z)}\frac{\delta V}{\delta \beta_j^{[\pm]}(\omega,\mathbf{q},z)}\right),
\end{eqnarray}
where the summation is over both directions and both values of $j$. Choosing $U=\beta_j^{[\pm]}(\omega,\mathbf{q},z)$, $V=\beta_l^{[\pm]*}(\omega',\mathbf{q}',z)$ or $V=\beta_l^{[\mp]*}(\omega',\mathbf{q}',z)$, we arrive at the canonical Poisson brackets in $\omega\mathbf{q}$ space:
\begin{eqnarray}\nonumber
 \left\{\beta_j^{[\pm]}(\omega,\mathbf{q},z),\beta_l^{[\pm]*}(\omega',\mathbf{q}',z)\right\}' &=& -\frac{i}\hbar\delta_{jl}\delta(\omega-\omega')\delta(\mathbf{q}-\mathbf{q}'),\\\label{PoissonPrimeBeta}
 \left\{\beta_j^{[\pm]}(\omega,\mathbf{q},z),\beta_l^{[\mp]*}(\omega',\mathbf{q}',z)\right\}' &=& 0.
\end{eqnarray}
Comparing these expressions with Eq.~(\ref{Poissonbeta}), we conclude that in the absence of sources the spatial Poisson bracket coincides with the conventional, temporal one.

In the general case of fields with sources, we can rewrite the equations for the normal variables, Eqs. (\ref{beta1diff}) and (\ref{beta2diff}), in a spatial Hamiltonian form
\begin{equation}\label{Hamilteqbeta}
\frac{\partial}{\partial z}\beta^{[\pm]}_j(\omega,\mathbf{q},z) = \left\{ \beta^{[\pm]}_j(\omega,\mathbf{q},z),G \right\}',
\end{equation}
where $G=G_F^{[+]}+G_F^{[-]}+G_I^{[+]}+G_I^{[-]}$ is the generator of spatial evolution, being a sum of four terms. The first two terms are the generators for the right- and left-propagating parts of the field
\begin{equation}\label{GF}
G_F^{[\pm]} =  \mp\hbar\sum_{j=1}^2\int d\mathbf{q}\int_{cq}^{+\infty}d\omega K_z\beta^{[\pm]}_j\beta^{[\pm]*}_j,
\end{equation}
while the last two terms describe the interaction with the sources
\begin{eqnarray}\nonumber
G_I^{[\pm]} &=&  -\hbar\int d\mathbf{q}\int_{cq}^{+\infty}d\omega\left[ \frac{\beta^{[\pm]*}_1K_z}{\epsilon_0c\mathcal{N}}(k_x\mathcal{J}_x+k_y\mathcal{J}_y \right. \\\label{GI}
 &\mp& \left.\frac{q^2}{K_z} \mathcal{J}_z  )\pm \frac{\beta^{[\pm]*}_2\omega}{\epsilon_0c^2\mathcal{N}}\left( k_x\mathcal{J}_y-k_y\mathcal{J}_x\right)\right]+ c.c.
\end{eqnarray}

Here, as in Sec.~\ref{sec:indep}, we assume that the currents and charges are fixed and do not depend on dynamical variables. The physical meaning of the function $G$ will become clear below, upon its expression through the fields. 

\subsection{Alternative Hamiltonian} 

Similar to the consideration of the temporal evolution in Sec.~\ref{sec:AltH}, we could construct alternative Hamiltonian formulations for the equations of motion, replacing all or some of the complex normal variables by their complex conjugates, e.g., defining $\tilde\beta^{[-]}_j(\omega,\mathbf{q},z)=\beta^{[-]*}_j(\omega,\mathbf{q},z)$. Subsequently, an alternative Poisson bracket could be defined with respect to the new variables and an alternative spatial Hamiltonian. Such an approach would inevitably result in appearance of the conjugates of the complex normal variables in the positive-frequency parts of the fields, determined by Eqs.~(\ref{Epm}) and (\ref{Bpm}). At the quantization stage, this would bring photon creation operators into the positive-frequency parts of the fields, which would be incompatible with the formalism of Glauber's theory of optical coherence. For this reason, we extend the validity of the Postulate on the positive-frequency part, formulated in Sec.~\ref{sec:AltH} to the consideration of spatial evolution, and reject all alternative Hamiltonian formulations of the equations of motion.

\subsection{Expressing the Hamiltonian via the fields}

As we have seen in Sec.~\ref{sec:Time-H}, the generator of temporal evolution is given by the full energy of the field. Now, we are interested in finding the corresponding physical quantity in the case of spatial evolution. Let us first establish the general requirements it should satisfy.

The energy of the field is obtained by integrating the energy density $T^{00}(\mathbf{r},t)$ over the three spatial coordinates. The result of such an integration is formally dependent on time. In the case of a time-invariant (conservative) system the full energy is constant. However, for a nonconservative system, which may be the case for fixed sources considered here, the field energy is time dependent. In this case the Heisenberg equation, Eq.~(\ref{Heisenberg}), requires a special technique, known as time ordering, for its solution.   

In a similar manner, the generator of spatial evolution $G$ can be $z$ dependent for a spatially noninvariant system. Note that its parts, defined by Eqs.~(\ref{GF}) and (\ref{GI}), depend on $z$ via the variables and the sources. However, it cannot depend on $t$, $x$, or $y$, being a functional of normal variables in $\omega\mathbf{q}$ space. 

In addition, the generator should be \emph{local} in $\omega\mathbf{q}$ space, i.e., every variable at point $\omega\mathbf{q}$ should be multiplied by another variable or source at the same point. Only such a local function produces a local equation of motion via the Poisson bracket. We recall that Maxwell's equations are local in $\omega\mathbf{q}$ space, as was established in Sec.~\ref{sec:Space-Maxwell}.

Thus, we formulate two ``principal requirements'' for the generator of spatial evolution in the $z$ direction: locality in $\omega\mathbf{q}$ space and independence of $t$, $x$, and $y$.

Below, we derive the expressions for the energy, momentum, and some other physical quantities of the field in  $\omega\mathbf{q}$ space with the aim of finding a function similar to $G_F=G_F^{[+]}+G_F^{[-]}$ for a field with sources or, at least, without them. At this step we do not consider the part of the generator describing the interaction with the sources, but we take into account that the presence of the latter influences the $z$ dependence of the normal variables, destroying their simple harmonic oscillation described by Eq.~(\ref{betafree}).

\subsubsection{Energy in $\omega\mathbf{q}$ space}

If we substitute the $\omega\mathbf{q}$-space representations of the electric and magnetic fields, found in Sec.~\ref{sec:Space-fields}, into the  energy density $T^{00}(\mathbf{r},t)$, defined by Eqs.~(\ref{T00F}) and (\ref{T00I}), and integrate over space, we obtain a complicated expression, including products of normal variables taken at different frequencies. In other words, energy is \emph{nonlocal} in $\omega\mathbf{q}$ space in striking contrast to the generator $G_F$, defined by Eq.~(\ref{GF}). A \emph{local} expression for the field energy is possible only in the case of a free field without sources, where we can use the solution, Eq.~(\ref{betafree}), giving us a delta function as a result of integration over $z$. In this case the field energy can be written as $H_f=H_f^{[+]}+H_f^{[-]}$, where
\begin{eqnarray}\nonumber
H_f^{[\pm]} &=& \frac{\epsilon_0}{2}\int \left(\left[\mathbf{E}^{[\pm]}_\mathrm{full}(\mathbf{r},t)\right]^2+c^2\left[\mathbf{B}^{[\pm]}_\mathrm{full}(\mathbf{r},t)\right]^2\right)d^3r\\\label{Hf}
&=& \hbar\int d\mathbf{q}\int_{cq}^\infty  d\omega\omega\sum_{j=1}^2\beta^{[\pm]}_j \beta^{[\pm]*}_j 
\end{eqnarray}
is the energy of the right- or left-propagating part of the field. Here, $\mathbf{E}^{[\pm]}_\mathrm{full}=\mathbf{E}^{[\pm]}+\mathbf{E}^{[\pm]*}$ is the full right- or left-propagating field, including both the positive- and negative-frequency parts, with a similar expression for $\mathbf{B}^{[\pm]}_\mathrm{full}$. Formally, Eq.~(\ref{T00F}) leads also to the appearance of cross-terms including the products $\beta^{[+]}_j\beta^{[-]}_j$ and their conjugates. However, they have opposite signs in the electric and magnetic parts of the full energy and cancel out.

The free-field energy $H_f$ is local and time independent, and thus satisfies the principal requirements, formulated above. However it does not coincide with the functional form of $G_F$ even in the absence of sources, Eq.~(\ref{GF}). 

\subsubsection{Momentum in $\omega\mathbf{q}$ space \label{sec:Momentum}}

Components of the field momentum $\mathbf{P}=(P_x,P_y,P_z)$ are obtained by integrating the momentum density 
\begin{equation}\label{T0i}
\frac1c T^{0i}(\mathbf{r},t) = \epsilon_0 \left[\mathbf{E}(\mathbf{r},t)\times\mathbf{B}(\mathbf{r},t)\right]\mathbf{n}_i
\end{equation}
over the entire space \cite{Landau-LifshitzII,Jackson}. Here the index $i$ takes values $1,2,3$, corresponding to $x,y,z$ respectively. Substituting the $\omega\mathbf{q}$-space representations of the electric and magnetic fields and performing the integration, we see that, similar to the energy, the field momentum is  time dependent in general and \emph{nonlocal} in $\omega\mathbf{q}$ space. A \emph{local} form is obtained again in the absence of sources, when the field momentum can be written as a sum of momenta of the right- and left-propagating fields, $\mathbf{P}_f=\mathbf{P}_f^{[+]}+\mathbf{P}_f^{[-]}$, where
\begin{eqnarray}\nonumber
\mathbf{P}_f^{[\pm]} &=& \epsilon_0 \int \mathbf{E}^{[\pm]}_\mathrm{full}(\mathbf{r},t)\times\mathbf{B}^{[\pm]}_\mathrm{full}(\mathbf{r},t)d^3r\\\label{Pf}
&=& \hbar\int d\mathbf{q}\int_{cq}^\infty  d\omega\mathbf{K}^{[\pm]}\sum_{j=1}^2\beta^{[\pm]}_j \beta^{[\pm]*}_j 
\end{eqnarray}
with $\mathbf{K}^{[\pm]}=(k_x,k_y,\pm K_z)$. The cross-terms including the products $\beta^{[+]}_j\beta^{[-]}_j$ and their conjugates are again zero identically, as it was for the energy.

The free-field momentum $\mathbf{P}_f$ is local and does not depend on $t$ or $z$. Its $z$ component, taken with the opposite sign, coincides with the generator of spatial evolution $G_F$, Eq.~(\ref{GF}). We conclude that, in the case of a free field, $G_F=-P_z$. However, in the presence of sources this equality does not hold in general.

\subsubsection{Transferred energy in $\omega\mathbf{q}$ space}

We have seen in the previous two sections that, by integrating a quadratic form of the fields over the space, we obtain a quantity which is time dependent and nonlocal in $\omega\mathbf{q}$ space in general. From the mathematical viewpoint, it would be more reasonable to integrate a quadratic form of the fields over $x$, $y$, and $t$, which would give a quantity dependent in general on $z$ and local in $\omega\mathbf{q}$ space, and thus satisfying automatically the two principal requirements for the generator of spatial evolution. From the physical viewpoint, the expression under such an integral should be a flux density of some quantity in the $z$ direction, so that the integral gives the amount of this quantity, transferred through the $(x,y)$ plane at point $z$ during the entire evolution time, from $t=-\infty$ to $+\infty$. This leads us to considering the elements $T^{3\mu}$, $\mu=0,1,2,3$, of the energy-momentum tensor, representing the flux density of the field energy and momentum in the $z$ direction \cite{Landau-LifshitzII,Jackson}.

In this section we consider the energy flux density $cT^{30}$. In accordance with its definition \cite{Landau-LifshitzII,Jackson}, we call the quantity
\begin{equation}
  S(z) = c\int T^{30}(\mathbf{r},t) dxdydt 
\end{equation}
the ``transferred energy'' through the point $z$. Due to the symmetry of the energy-momentum tensor in  vacuum, $T^{30}$ is equal to $T^{03}$ defined by Eq.~(\ref{T0i}). Substituting the $\omega\mathbf{q}$-space representations of the electric and magnetic fields and performing the integration, we find that the transferred energy can be written as $S=S^{[+]}+S^{[-]}$, where
\begin{eqnarray}\nonumber
S^{[\pm]} &=& \epsilon_0c^2 \int \left[\mathbf{E}^{[\pm]}_\mathrm{full}(\mathbf{r},t)\times\mathbf{B}^{[\pm]}_\mathrm{full}(\mathbf{r},t)\right]\mathbf{n}_zdxdydt\\\label{Spm}
&=& \pm\hbar\int d\mathbf{q}\int_{cq}^\infty  d\omega\omega\sum_{j=1}^2\beta^{[\pm]}_j \beta^{[\pm]*}_j, 
\end{eqnarray}
and the cross-terms including the products $\beta^{[+]}_j\beta^{[-]*}_j$ and their conjugates are zero identically. In the absence of sources, both $S^{[+]}$ and $S^{[-]}$ are $z$ independent and  Eq.~(\ref{Spm}) admits a simple interpretation: $S^{[+]}$ is the energy of the right-propagating field, crossing any plane $z=z_0$ in the positive direction, while $-S^{[-]}$ is the energy of the left-propagating field, crossing the same plane in the negative direction. According to the general definition of flux, the energy of the field crossing a given surface in the positive or negative direction is taken with the positive or negative sign, respectively. For this reason, $S$ is a difference of the energies of the right- and left-propagating fields.

The transferred energy $S$ satisfies the principal requirements for a generator of spatial evolution. In a spatially one-dimensional model, where $\mathbf{q}=0$ and $K_z=|\omega|/c$, we find $G_F=-S/c$, which corresponds to the result of Abram \cite{Abram87}. However, in the general four-dimensional model considered here, the transferred energy $S$  does not equal the expression for $G_F$.

\subsubsection{Transferred momentum in $\omega\mathbf{q}$ space}

We consider now the element $T^{33}$ of the energy-momentum tensor, which determines the flux density of the $z$ component of the field momentum \cite{Landau-LifshitzII, Jackson}, and call the quantity
\begin{equation}
  M(z) = \int T^{33}(\mathbf{r},t) dxdydt 
\end{equation}
the ``transferred momentum'' through the point $z$. We mean always the $z$ component of momentum, though we do not write it for brevity. A similar quantity was considered in Ref.~\cite{Ben-Aryeh91} for a one-directional propagation of a spatially one-dimensional field.

The momentum flux density can be written as \cite{Landau-LifshitzII, Jackson}
\begin{equation}\label{T33}
T^{33}(\mathbf{r},t) = \frac{\epsilon_0}2 \left[\mathbf{E}(\mathbf{r},t)\otimes\mathbf{E}(\mathbf{r},t)+c^2\mathbf{B}(\mathbf{r},t)\otimes\mathbf{B}(\mathbf{r},t)\right],
\end{equation}
where we have introduced a pseudo-Euclidean scalar product of two vectors $\mathbf{V}$ and $\mathbf{W}$: $\mathbf{V}\otimes\mathbf{W}=V_xW_x+V_yW_y-V_zW_z$. Substituting the $\omega\mathbf{q}$-space representations of the electric and magnetic fields, performing the integration, and taking into account the properties of the basis vectors,
\begin{eqnarray}\nonumber
\mathbf{m}_1(\omega,\mathbf{q})\otimes\mathbf{m}_1(\omega,\mathbf{q}) &=& \cos^2\theta(\omega,\mathbf{q})-\sin^2\theta(\omega,\mathbf{q}),\\\nonumber
\mathbf{m}_1(\omega,\mathbf{q})\otimes\mathbf{m}_1(\omega,-\mathbf{q}) &=& -1,\\
\mathbf{m}_2(\omega,\mathbf{q})\otimes\mathbf{m}_2(\mathbf{q}) &=& 1,\\\nonumber
\mathbf{m}_1(\omega,\mathbf{q})\otimes\mathbf{m}_2(\mathbf{q}) &=& 0,
\end{eqnarray}
we find that the transferred momentum can be written as $M=M^{[+]}+M^{[-]}$, where
\begin{eqnarray}\nonumber
M^{[\pm]} &=& \frac{\epsilon_0}{2}\int \left(\left[\mathbf{E}^{[\pm]}_\mathrm{full}(\mathbf{r},t)\right]^{\otimes2}+c^2\left[\mathbf{B}^{[\pm]}_\mathrm{full}(\mathbf{r},t)\right]^{\otimes2}\right)d\mathbf{x}dt\\\label{Mpm}
&=& \hbar\int d\mathbf{q}\int_{cq}^\infty  d\omega K_z\sum_{j=1}^2\beta^{[\pm]}_j \beta^{[\pm]*}_j,
\end{eqnarray}
and we have used the notation $\mathbf{V}^{\otimes2}=\mathbf{V}\otimes\mathbf{V}$. The cross-terms, including the products of variables of the right- and left-propagating fields, cancel out as in the three previously considered cases.

The transferred momentum is local in $\omega\mathbf{q}$ space and may depend on $z$ via the normal variables, but not on $x$, $y$, or $t$. Thus, it satisfies the principal requirements for the spatial Hamiltonian. Comparing Eq.~(\ref{Mpm}) to Eq.~(\ref{GF}), we conclude that  $G_F=-(M^{[+]}-M^{[-]})$. 

\begin{figure}[h]
\centering
\includegraphics[width=0.49\columnwidth]{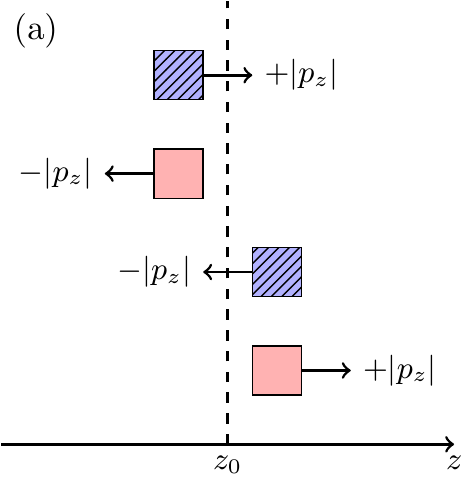}
\includegraphics[width=0.49\columnwidth]{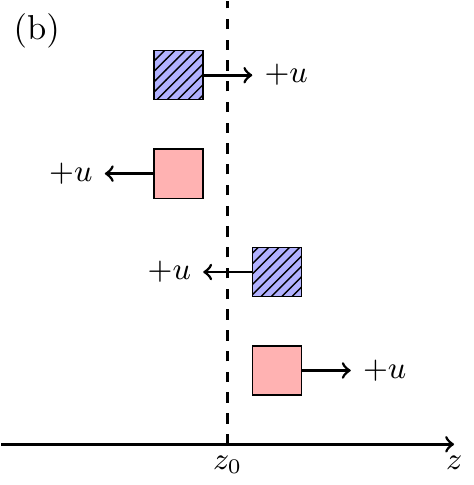}
\caption{Transfer of momentum (a) and energy (b) of the electromagnetic field through the plane $z=z_0$. Elementary volumes contributing to the transferred energy or momentum are shaded and marked in blue; those not contributing are not shaded and marked in red. Arrows show the direction of propagation of an elementary volume, while their labels show the amount of carried quantity, energy $u$, or momentum $p_z$.\label{fig:Elements}}
\end{figure}

To clarify the physical meaning of the  quantity $M^{[+]}-M^{[-]}$ we make the following observation. For any given point $z_0$, the transferred energy $S(z_0)$ may be positive or negative, depending on which part of the field, right- or left-propagating, has a higher energy. On the other hand, the transferred momentum $M(z_0)$ is always positive. This happens because an elementary volume with a positive momentum $+|p_z|$ crosses the plane $z=z_0$ in the positive direction, while an elementary volume with a negative momentum $-|p_z|$ crosses this plane in the negative direction [see Fig.~\ref{fig:Elements}(a)]. Since the sign of the transferred momentum depends on the direction in which the surface is crossed, the contributions of both elementary volumes to the transferred momentum are positive. In contrast to momentum, the carried energy $u$ is positive for all elementary volumes independently of their directions [see Fig.~\ref{fig:Elements}(b)], and the transferred energy is positive for the right-propagating field and negative for the left-propagating one. If, similarly, we assign to each elementary volume the \emph{modulus of its momentum} $|p_z|$, which is always positive, then the total transferred modulus of momentum of the right-propagating field will be $M^{[+]}(z_0)$, while that of the left-propagating field will be $-M^{[-]}(z_0)$. Finally, we have found that the quantity $M^{[+]}(z_0)-M^{[-]}(z_0)$ has a meaning of the \emph{transferred modulus of momentum} through the plane $z=z_0$. The spatial Hamiltonian of the field $G_F(z_0)$ is equal to this quantity taken with the opposite sign. This equality holds in the presence of sources, in contrast to the equality involving the field momentum $P_z$.

Our expression for the spatial evolution generator reduces to that of Ref.~\cite{Ben-Aryeh91} in the case of one-directional propagation of the field. It can be said that our approach combines the idea of Ben-Aryeh and Serulnik~\cite{Ben-Aryeh91}, namely, integrating $T^{33}$ over the transverse plane and time, and that of Abram \cite{Abram87}, namely, splitting the field into the left- and right-propagating parts and taking a difference of some quantities calculated separately for them.

\subsubsection{Interaction with sources}

We turn now to the part of the generator, which describes the interaction between the field and the sources. Equation (\ref{GI}) can be rewritten as
\begin{equation}
    G_I^{[\pm]}=\mp\int  \mathbf{J}(\mathbf{r},t)\left[\mathbf{A}^{[\pm]}(\mathbf{r},t)+\mathbf{A}^{[\pm]*}(\mathbf{r},t)\right] d\mathbf{x}dt,
\end{equation}
where 
\begin{eqnarray}\label{Apm}
\mathbf{A}^{[\pm]}(\mathbf{r},t) &=& \frac{1}c\int d\mathbf{q}\int_{cq}^\infty d\omega \sqrt{\frac{\hbar}{(2\pi)^3 2\epsilon_0 K_z}} \\\nonumber 
&\times& \left[\beta_{1}^{[\pm]}\mathbf{m}_1(\omega,\pm\mathbf{q}) +\beta_{2}^{[\pm]}\mathbf{m}_2(\mathbf{q})\right] e^{i\mathbf{qx}-i\omega t}
\end{eqnarray}
is a vector, which can be interpreted as the positive-frequency part of the vector potential of the right- or left-propagating field. In order to satisfy the relation $\mathbf{B}=\nabla\times\mathbf{A}$, we need to write the vector potential as a sum of six terms $\mathbf{A}=\mathbf{A}^{[+]}+\mathbf{A}^{[-]}+\mathbf{A}^{[s]}+c.c.$, where
\begin{equation}\label{As}
\mathbf{A}^{[s]}(\mathbf{r},t) = -\int d\mathbf{q}\int_{cq}^\infty d\omega \frac{\mathcal{J}_z(\omega,\mathbf{q},z)}{(2\pi)^{3/2}\epsilon_0\omega^2}\mathbf{n}_z e^{i\mathbf{qx}-i\omega t}
\end{equation}
is a part determined by the sources. From the $\omega\mathbf{q}$-representation of the electric field, given by Eqs.~(\ref{Ert}), (\ref{Epm}) and (\ref{Es}), we find $\mathbf{E}=-\dot{\mathbf{A}}$, which allows us to let the scalar potential vanish, $\Phi(\mathbf{r},t)=0$, meaning the \emph{temporal gauge} \cite{Jackson02} for the field. In contrast to the Coulomb gauge, employed in Sec.~\ref{sec:Time}, in the temporal gauge the divergence of the vector potential is nonzero:
\begin{equation}\label{divA}
\nabla\mathbf{A}(\mathbf{r},t) = -i\int d\mathbf{q}\int_{-\infty}^{+\infty} d\omega \frac{\mathcal{R}(\omega,\mathbf{q},z)}{(2\pi)^{3/2}\epsilon_0\omega} e^{i\mathbf{qx}-i\omega t}.
\end{equation}
The Coulomb gauge may be obtained in a standard way \cite{Cohen-Tannoudji89,Landau-LifshitzII}, by adding to the vector potential the gradient of a gauge function $F(\mathbf{r},t)$, such that $\Delta F=-\nabla\mathbf{A}$, where $\Delta$ stands for the Laplacian. The resulting vector potential $\mathbf{A}'=\mathbf{A}+\nabla F$ is transverse. Correspondingly, the scalar potential becomes $\Phi'=-\dot{F}$ and satisfies $\Delta\Phi'=\nabla\dot{\mathbf{A}}$. Differentiating both sides of Eq.~(\ref{divA}) by time we find $\Delta\Phi'=-\rho/\epsilon_0$, as expected for the Coulomb gauge.

\subsubsection{Total spatial Hamiltonian}

Similar to the total energy density, playing the key role in the description of the temporal evolution, we introduce the total momentum flux density of field and sources $T^{33}_{FS}=T^{33}+T^{33}_I+T^{33}_S$, where $T^{33}_I=\mathbf{J}(\mathbf{r},t)\mathbf{A}(\mathbf{r},t)$ is the part describing the interaction, while $T^{33}_S$ is the momentum flux density of the sources, which can be disregarded for particles independent of the field, since it does not contain dynamical variables. Note that the sign of $T^{33}_I$ is opposite to that of $T^{00}_I$, Eq.~(\ref{T00I}). This sign is determined by the corresponding element of the metric tensor,  $g^{00}=1$ or $g^{33}=-1$, in the definition of the energy-momentum tensor via the Lagrangian density \cite{Landau-LifshitzII}.

To write the total spatial Hamiltonian through the fields, we do the following. First, we split the fields into the right- and left-propagating parts. For the magnetic field this means $\mathbf{B}=\mathbf{B}^{[+]}_\mathrm{full}+\mathbf{B}^{[-]}_\mathrm{full}$, which is mathematically equivalent to the decomposition into the positive- and negative-frequency parts \cite{Glauber63a,Mandel-Wolf}. For the electric field, we need first to subtract the field directly determined by the sources and then do the decomposition: $\mathbf{E}-
\mathbf{E}^{[s]}-\mathbf{E}^{[s]*}=\mathbf{E}^{[+]}_\mathrm{full}+\mathbf{E}^{[-]}_\mathrm{full}$. The same procedure is done for the vector potential in the temporal gauge: $\mathbf{A}-
\mathbf{A}^{[s]}-\mathbf{A}^{[s]*}=\mathbf{A}^{[+]}_\mathrm{full}+\mathbf{A}^{[-]}_\mathrm{full}$. For the latter two fields this procedure is similar to selecting the transverse part in the description of the temporal evolution.

Second, we calculate the total momentum flux density for each part of the field as
\begin{eqnarray}\nonumber
T^{33[\pm]}_{FS} &=& \frac{\epsilon_0}{2} \left[\mathbf{E}^{[\pm]}_\mathrm{full}(\mathbf{r},t)\right]^{\otimes2}+\frac{\epsilon_0c^2}{2}\left[\mathbf{B}^{[\pm]}_\mathrm{full}(\mathbf{r},t)\right]^{\otimes2}\\\label{TFSpm}
&+& \mathbf{J}(\mathbf{r},t)\mathbf{A}^{[\pm]}_\mathrm{full}(\mathbf{r},t).
\end{eqnarray}
In the case of particles depending on the field, we need to complement this expression by the momentum flux density of the particles. However, this lies outside the scope of the present paper. 

Finally, we write the spatial Hamiltonian (the generator of spatial evolution) as 
\begin{equation}\label{G}
G = - \int \left(T^{33[+]}_{FS}-T^{33[-]}_{FS}\right) d\mathbf{x}dt.
\end{equation}

\subsection{Spatial Lagrangian} 

To find the spatial Lagrangian, we introduce real normal variables $X_j^{[\pm]}(\omega,\mathbf{q},z)$ and $Y_j^{[\pm]}(\omega,\mathbf{q},z)$ in a way similar to that of Sec.~\ref{sec:Time-normal}:  
\begin{equation}\label{X}
\beta_j^{[\pm]} = \sqrt{\frac{\epsilon_0}{2\hbar K_z}}\left(K_z X_j^{[\pm]}+\frac{i}{\epsilon_0}Y_j^{[\pm]}\right).
\end{equation}
In terms of these variables we rewrite the equations of motion, Eqs.~(\ref{beta1diff}) and (\ref{beta2diff}), as
\begin{eqnarray}\nonumber
\frac\partial{\partial z}X_j^{[\pm]} &=& \mp\frac1{\epsilon_0} Y_j^{[\pm]} \mp \frac1{\epsilon_0cK_z}\Ima\mathcal{J}_j^{[\pm]},\\\label{Xdiff}
\frac\partial{\partial z}Y_j^{[\pm]} &=& \pm\epsilon_0K_z^2 X_j^{[\pm]} \pm \frac1c\Rea\mathcal{J}_j^{[\pm]},
\end{eqnarray}
where
\begin{eqnarray}\nonumber
\mathcal{J}_1^{[\pm]}&=& -\mathcal{J}_z\sin\theta \pm\left(\mathcal{J}_x\cos\varphi+\mathcal{J}_y\sin\varphi\right)\cos\theta,  \\\label{J1}
\mathcal{J}_2^{[\pm]} &=& \mathcal{J}_y\cos\varphi-\mathcal{J}_x\sin\varphi
\end{eqnarray}
are projections of the vector $\mathbf{\mathcal{J}}(\omega,\mathbf{q},z)$ on the vectors $\mathbf{m}_1(\omega,\pm\mathbf{q})$ and $\mathbf{m}_2(\mathbf{q})$ respectively. 

Excluding $Y_j^{[\pm]}$ from Eq.~(\ref{Xdiff}), we obtain a second-order equation for $X_j^{[\pm]}$, which can be regarded as a spatial Lagrange equation with the Lagrangian density (see Appendix~\ref{appendix:Lagrangian})
\begin{eqnarray}\label{LX}
\mathcal{L}_X^{[\pm]} &=& \frac{\epsilon_0}2\sum_{j=1}^2 \left[\left(\partial_zX_j^{[\pm]}\right)^2-K_z^2X_j^{[\pm]2}\right.\\\nonumber
&-& \left.\frac2{c\epsilon_0} X_j^{[\pm]}\Rea\left(J_j^{[\pm]}\pm\frac1{iK_z}\partial_z J_j^{[\pm]}\right)\right].
\end{eqnarray}

Excluding $X_j^{[\pm]}$ from Eq.~(\ref{Xdiff}), we obtain a second-order equation for $Y_j^{[\pm]}$, leading us to the Lagrangian density
\begin{eqnarray}\label{LY}
\mathcal{L}_Y^{[\pm]} &=& \frac1{2\epsilon_0K_z^2}\sum_{j=1}^2 \left[\left(\partial_zY_j^{[\pm]}\right)^2-K_z^2Y_j^{[\pm]2}\right.\\\nonumber
&-& \left.\frac{2K_z}{c} Y_j^{[\pm]}\Ima\left(J_j^{[\pm]}\pm\frac1{iK_z}\partial_z J_j^{[\pm]}\right)\right].
\end{eqnarray}

In the direct space, two Lagrangians can be constructed as 
\begin{eqnarray}\label{Lpm}
L^{[\pm]} &=& \frac12\int d\mathbf{x}dt \left[\epsilon_0c^2\left(\partial_z\mathbf{A}_\mathrm{full}^{[\pm]}\right)^2\right.\\\nonumber 
&+&\epsilon_0c^2 \mathbf{A}_\mathrm{full}^{[\pm]}\Delta' \mathbf{A}_\mathrm{full}^{[\pm]}
- \left.\left(\mathbf{J}+\mathbf{W}^{[\pm]}\right) \mathbf{A}_\mathrm{full}^{[\pm]}\right],
\end{eqnarray}
where $\Delta'=c^{-2}\partial_t^2-\partial_x^2-\partial_y^2$ is a pseudo-Euclidean Laplacian, while
\begin{eqnarray}\nonumber
\mathbf{W}^{[\pm]}(\mathbf{r},t) &=& \frac{\pm 1}{(2\pi)^\frac32} \int d\mathbf{q}\int_{cq}^{\infty}d\omega \frac{\partial_z\mathbf{\mathcal{J}}(\omega,\mathbf{q},z)}{iK_z} e^{i\mathbf{qx} -i\omega t} \\\label{W}
&+& c.c.
\end{eqnarray}

Substituting the vector potential from Eq.~(\ref{Apm}) into Eq.~(\ref{Lpm}), we obtain
\begin{equation}\label{Lfinal}
L^{[\pm]} = \frac12\int d\mathbf{q}\int_{cq}^{\infty}d\omega \left(\mathcal{L}_X^{[\pm]}+\mathcal{L}_Y^{[\pm]}\right),
\end{equation}
which shows that the full spatial Lagrangian $L_S=L^{[+]}+L^{[-]}$ is redundant: it gives correct equations of motion when regarded either as a functional of the coordinates $X_j^{[\pm]}$ and velocities $\partial_zX_j^{[\pm]}$ or as a functional of the coordinates $Y_j^{[\pm]}$ and velocities $\partial_zY_j^{[\pm]}$. As a consequence, $L_S$ can be regarded as a functional of the complex generalized coordinate $\beta_j^{[\pm]}(\omega,\mathbf{q},z)$, the generalized velocity $\partial_z\beta_j^{[\pm]}(\omega,\mathbf{q},z)$, and their complex conjugates, and the corresponding Lagrange equations give the correct equations of motion.

\subsection{Field quantization} 

Having obtained a classical Hamiltonian formulation of the equations of motion, we proceed to the field quantization in a standard way, as it was done in Sec.~\ref{sec:Time-Quant}. The complex normal variables $\beta_j^{[\pm]}(\omega,\mathbf{q},z)$ and $\beta_j^{[\pm]*}(\omega,\mathbf{q},z)$ are replaced by operators $b_j^{[\pm]}(\omega,\mathbf{q},z)$ and $b_j^{[\pm]\dagger}(\omega,\mathbf{q},z)$, having the meanings of annihilation and creation operators, respectively, for a photon at point $z$ with frequency $\omega$, transverse wave vector $\mathbf{q}$, and polarization $j$ and propagating in the positive or negative direction with respect to the $z$ axis. The canonical commutators for these operators are given by the Poisson brackets, Eq.~(\ref{PoissonPrimeBeta}), multiplied by $i\hbar$: 
\begin{eqnarray}\nonumber
 \left[b_j^{[\pm]}(\omega,\mathbf{q},z),b_l^{[\pm]\dagger}(\omega',\mathbf{q}',z)\right] &=& \delta_{jl}\delta(\omega-\omega')\delta(\mathbf{q}-\mathbf{q}'),\\\label{Commb}
 \left[b_j^{[\pm]}(\omega,\mathbf{q},z),b_l^{[\mp]\dagger}(\omega',\mathbf{q}',z)\right] &=& 0.
\end{eqnarray}

The fields also become operators, retaining their expressions through the normal variables. The spatial Hamiltonian is given by Eq.~(\ref{G}), determining the dependence of $G$ on the photon annihilation and creation operators $b_j^{[\pm]}(\omega,\mathbf{q},z)$ and $b_j^{[\pm]\dagger}(\omega,\mathbf{q},z)$.

The Hamilton equation~(\ref{Hamilteqbeta}) is replaced by the spatial Heisenberg equation
\begin{equation}\label{SpatHeisenberg}
i\hbar\frac{\partial}{\partial z}b_j^{[\pm]}(\omega,\mathbf{q},z) = \left[ b_j^{[\pm]}(\omega,\mathbf{q},z),G \right].
\end{equation}

In the absence of sources, considering propagation in one direction, and disregarding the transverse dimensions, we obtain $G=-P$, where $P$ is the field momentum, which reproduces the formalism of Shen \cite{Shen67}, widely used for treating quantum optical problems connected to propagation of quantized fields through linear \cite{Horoshko18JOSAB} and nonlinear media \cite{Huttner90,Barral20a,Barral20b,Linares08}. As it was shown in Sec.~\ref{sec:Momentum}, the relation $G=-P_z$ remains valid in four-dimensional models including propagation of a free field in both directions~\cite{Perina95,Perina00,Horoshko19,LaVolpe21,Horoshko12epjd}. However, for a field with sources, the form of the generator of spatial evolution is given by the quantum version of Eq.~(\ref{G}).

\section{Conclusion \label{sec:Conclusion}}

In this paper, the problem of building a quantum Hamiltonian formalism for the spatial evolution of the electromagnetic field was approached by an \emph{inductive} derivation from Maxwell's equations. We have limited our consideration to the field with fixed sources, obeying some rather general restrictions, and found, that for a unique determination of the field Hamiltonian, one needs an additional rule, formulated in a form of Postulate on the positive-frequency part. Basing our approach on this postulate, we have derived unambiguously the generator of the spatial evolution and clarified its physical meaning: the modulus of momentum transferred through the transverse plane at a given point $z$. 

We hope that this approach can be extended to more complicated configurations including dielectric media and nonlinear phenomena, and will serve as a starting point for consideration of the spatial evolution of electromagnetic field in the optical and other spectral ranges, where an explicit form of the evolution generator is required.   

\begin{acknowledgments}
This work was supported by the Habilitation research support program of the National Academy of Sciences of Belarus and by Belarusian Republican Foundation for Fundamental Research under grant F21TURG-003.
\end{acknowledgments}

\appendix
\section{Complex Poisson bracket \label{appendix:Poisson}}

First, we consider one mode of an electromagnetic field with frequency $ck$, which is described by a generalized coordinate $Q$ and a generalized momentum $P$. The Poisson bracket for any two functions $\bar{U}(Q,P)$ and $\bar{V}(Q,P)$ is defined as \cite{Goldstein80}
\begin{equation}\label{Poisson-qp}
\left\{ \bar{U},\bar{V} \right\} = \frac{\partial\bar{U}}{\partial Q} \frac{\partial\bar{V}}{\partial P} - \frac{\partial\bar{U}}{\partial P} \frac{\partial\bar{V}}{\partial Q}.
\end{equation}

We introduce two linear combinations of the generalized coordinate and momentum,
\begin{equation}
\alpha = \sqrt{\frac{\epsilon_0}{2\hbar ck}}\left(ck Q + \frac{i}{\epsilon_0}P\right)
\end{equation}
and its complex conjugate $\alpha^*$, which are considered as new independent variables. Since
\begin{eqnarray}
\frac\partial{\partial Q} &=& \sqrt{\frac{\epsilon_0ck}{2\hbar}}\left(\frac\partial{\partial\alpha}+\frac\partial{\partial\alpha^*}\right),\\
\frac\partial{\partial P} &=&  \frac{i}{\sqrt{2\hbar ck\epsilon_0}} \left(\frac\partial{\partial\alpha} -\frac\partial{\partial\alpha^*}\right),
\end{eqnarray}
we rewrite the Poisson bracket, Eq. (\ref{Poisson-qp}), for any two functions $U(\alpha,\alpha^*)=\bar{U}(q,p)$ and $V(\alpha,\alpha^*)=\bar{V}(q,p)$ as
\begin{equation}\label{Poisson-alpha}
\left\{ U,V \right\} = -\frac{i}\hbar\left(\frac{\partial U}{\partial \alpha}\frac{\partial V}{\partial \alpha^*} - \frac{\partial U}{\partial \alpha^*}\frac{\partial V}{\partial \alpha}\right).
\end{equation}
Taking $U=V^*=\alpha$, we obtain $\left\{ \alpha,\alpha^* \right\} = -i/\hbar$. 

For a discrete set of normal variables, typical for considering the field with periodic boundary conditions \cite{Mandel-Wolf}, the Poisson bracket is defined for any two functions $U$ and $V$ of the normal variables $\alpha_1(\mathbf{k})$ and $\alpha_2(\mathbf{k})$ and their complex conjugates as
\begin{equation}\label{Poisson-discrete}
\left\{ U,V \right\} = -\frac{i}\hbar\sum_{\mathbf{k}}\sum_{j=1}^2\left(\frac{\partial U}{\partial \alpha_j(\mathbf{k})}\frac{\partial V}{\partial \alpha_j^*(\mathbf{k})} - \frac{\partial U}{\partial \alpha_j^*(\mathbf{k})}\frac{\partial V}{\partial \alpha_j(\mathbf{k})}\right),
\end{equation}
where the first sum is over all discrete values of $\mathbf{k}$.

Equation (\ref{Poisson}) is a generalization of the latter relation to the case of continuous dependence of the normal variables on the wave vector. It includes a functional derivative \cite{Barnett97}, which in the case of a linear functional 
\begin{equation}\label{Functional}
U = \int f(\mathbf{k})\alpha_j(\mathbf{k})d^3k,
\end{equation}
is simply 
\begin{equation}\label{FuncDer}
\frac{\delta U}{\delta\alpha_j(\mathbf{k})} = f(\mathbf{k}).
\end{equation}
All functionals of the field considered in the present paper are linear in a given normal variable, though the corresponding $f(\mathbf{k})$ may depend on other variables. Thus,  Eq.~(\ref{FuncDer}) can be used for calculating all functional derivatives. In particular, choosing $U_0=\alpha_m(\mathbf{k}_0)$, which we can rewrite as
\begin{equation}\label{alphaj}
U_0 = \int \delta(\mathbf{k}-\mathbf{k}_0)\alpha_m(\mathbf{k})d^3k,
\end{equation}
we arrive at 
\begin{equation}\label{FuncDer0}
\frac{\delta U_0}{\delta\alpha_j(\mathbf{k})} =\delta_{mj}\delta(\mathbf{k}-\mathbf{k}_0),
\end{equation}
which makes straightforward the derivation of  Eq.~(\ref{Poisson1}) from Eq.~(\ref{Poisson}).

\section{Lagrangian of a driven oscillator \label{appendix:Lagrangian}}

Let us consider the temporal evolution of a mechanical oscillator of mass $m$ on a spring with stiffness $\xi$ under the action of external forces $f(t)$ and $g(t)$, described by the Lagrangian
\begin{equation}\label{L0}
L = \frac{m\dot{x}^2}2 - \frac{\xi x^2}2 + f(t)x + g(t)\dot{x},
\end{equation}
which corresponds to Hamiltonian dynamics with the Hamilton equations for $x$ and its conjugate momentum $p=m\dot{x}+g(t)$:
\begin{eqnarray}\label{Hameq0}
\dot{x} &=& \frac{p}m - \frac{g(t)}m,\\\nonumber
\dot{p} &=& -\xi x + f(t).
\end{eqnarray}

Since the Lagrangian is defined up to a time derivative of an arbitrary function of time \cite{Landau-LifshitzI,Goldstein80}, the same dynamics is described by the Lagrangian
\begin{equation}\label{L1}
\tilde{L} = \frac{m\dot{x}^2}2 - \frac{\xi x^2}2 + f(t)x - \dot{g}(t)x,
\end{equation}
which corresponds to Hamiltonian dynamics for $x$ and its conjugate momentum $\tilde{p}=m\dot{x}$
\begin{eqnarray}
\dot{x} &=& \frac{\tilde p}m,\\\nonumber
\dot{\tilde p} &=& -\xi x + f(t) - \dot g(t).
\end{eqnarray}

We see that the addition of the full time derivative of the function $-g(t)x$ to the Lagrangian is equivalent to the canonical transformation $(x,p)\to(x,\tilde p)$.

Now, we consider Eq.~(\ref{Xdiff}) as the spatial analog of Eq.~(\ref{Hameq0}) having a similar structure. Identifying $x=X_j^{[\pm]}$, $p=\mp Y_j^{[\pm]}$, $m=\epsilon_0$, $\xi=\epsilon_0K_z^2$, and substituting these expressions into Eq.~(\ref{L1}), we obtain $\mathcal{L}_X^{[\pm]}$, Eq.~(\ref{LX}). 

Alternatively, identifying $x=Y_j^{[\pm]}$, $p=\pm X_j^{[\pm]}$, $m=1/\left(\epsilon_0K_z^2\right)$, $\xi=1/\epsilon_0$, and substituting these expressions into Eq.~(\ref{L1}), we obtain $\mathcal{L}_Y^{[\pm]}$, Eq.~(\ref{LY}).

\bibliography{SpatialEvolution}

\providecommand{\noopsort}[1]{}\providecommand{\singleletter}[1]{#1}%
\begin{thebibliography}{44}%
\makeatletter
\providecommand \@ifxundefined [1]{%
 \@ifx{#1\undefined}
}%
\providecommand \@ifnum [1]{%
 \ifnum #1\expandafter \@firstoftwo
 \else \expandafter \@secondoftwo
 \fi
}%
\providecommand \@ifx [1]{%
 \ifx #1\expandafter \@firstoftwo
 \else \expandafter \@secondoftwo
 \fi
}%
\providecommand \natexlab [1]{#1}%
\providecommand \enquote  [1]{``#1''}%
\providecommand \bibnamefont  [1]{#1}%
\providecommand \bibfnamefont [1]{#1}%
\providecommand \citenamefont [1]{#1}%
\providecommand \href@noop [0]{\@secondoftwo}%
\providecommand \href [0]{\begingroup \@sanitize@url \@href}%
\providecommand \@href[1]{\@@startlink{#1}\@@href}%
\providecommand \@@href[1]{\endgroup#1\@@endlink}%
\providecommand \@sanitize@url [0]{\catcode `\\12\catcode `\$12\catcode
  `\&12\catcode `\#12\catcode `\^12\catcode `\_12\catcode `\%12\relax}%
\providecommand \@@startlink[1]{}%
\providecommand \@@endlink[0]{}%
\providecommand \url  [0]{\begingroup\@sanitize@url \@url }%
\providecommand \@url [1]{\endgroup\@href {#1}{\urlprefix }}%
\providecommand \urlprefix  [0]{URL }%
\providecommand \Eprint [0]{\href }%
\providecommand \doibase [0]{https://doi.org/}%
\providecommand \selectlanguage [0]{\@gobble}%
\providecommand \bibinfo  [0]{\@secondoftwo}%
\providecommand \bibfield  [0]{\@secondoftwo}%
\providecommand \translation [1]{[#1]}%
\providecommand \BibitemOpen [0]{}%
\providecommand \bibitemStop [0]{}%
\providecommand \bibitemNoStop [0]{.\EOS\space}%
\providecommand \EOS [0]{\spacefactor3000\relax}%
\providecommand \BibitemShut  [1]{\csname bibitem#1\endcsname}%
\let\auto@bib@innerbib\@empty
\bibitem [{\citenamefont {Cohen-Tannoudji}\ \emph {et~al.}(1989)\citenamefont
  {Cohen-Tannoudji}, \citenamefont {Dupont-Roc},\ and\ \citenamefont
  {Grynberg}}]{Cohen-Tannoudji89}%
  \BibitemOpen
  \bibfield  {author} {\bibinfo {author} {\bibfnamefont {C.}~\bibnamefont
  {Cohen-Tannoudji}}, \bibinfo {author} {\bibfnamefont {J.}~\bibnamefont
  {Dupont-Roc}},\ and\ \bibinfo {author} {\bibfnamefont {G.}~\bibnamefont
  {Grynberg}},\ }\href@noop {} {\emph {\bibinfo {title} {Photons and Atoms:
  Introduction to Quantum Electrodynamics}}}\ (\bibinfo  {publisher} {Wiley},\
  \bibinfo {address} {New York},\ \bibinfo {year} {1989})\BibitemShut {NoStop}%
\bibitem [{\citenamefont {Bloembergen}(1965)}]{Bloembergen65}%
  \BibitemOpen
  \bibfield  {author} {\bibinfo {author} {\bibfnamefont {N.}~\bibnamefont
  {Bloembergen}},\ }\href@noop {} {\emph {\bibinfo {title} {Nonlinear
  optics}}}\ (\bibinfo  {publisher} {W. A. Benjamin},\ \bibinfo {address} {New
  York},\ \bibinfo {year} {1965})\BibitemShut {NoStop}%
\bibitem [{\citenamefont {Shen}(1967)}]{Shen67}%
  \BibitemOpen
  \bibfield  {author} {\bibinfo {author} {\bibfnamefont {Y.~R.}\ \bibnamefont
  {Shen}},\ }\bibfield  {title} {\bibinfo {title} {Quantum statistics of
  nonlinear optics},\ }\href {https://doi.org/10.1103/PhysRev.155.921}
  {\bibfield  {journal} {\bibinfo  {journal} {Phys. Rev.}\ }\textbf {\bibinfo
  {volume} {155}},\ \bibinfo {pages} {921} (\bibinfo {year}
  {1967})}\BibitemShut {NoStop}%
\bibitem [{\citenamefont {Caves}\ and\ \citenamefont {Crouch}(1987)}]{Caves87}%
  \BibitemOpen
  \bibfield  {author} {\bibinfo {author} {\bibfnamefont {C.~M.}\ \bibnamefont
  {Caves}}\ and\ \bibinfo {author} {\bibfnamefont {D.~D.}\ \bibnamefont
  {Crouch}},\ }\bibfield  {title} {\bibinfo {title} {Quantum wideband
  traveling-wave analysis of a degenerate parametric amplifier},\ }\href
  {https://doi.org/10.1364/JOSAB.4.001535} {\bibfield  {journal} {\bibinfo
  {journal} {J. Opt. Soc. Am. B}\ }\textbf {\bibinfo {volume} {4}},\ \bibinfo
  {pages} {1535} (\bibinfo {year} {1987})}\BibitemShut {NoStop}%
\bibitem [{\citenamefont {Kolobov}(1999)}]{Kolobov99}%
  \BibitemOpen
  \bibfield  {author} {\bibinfo {author} {\bibfnamefont {M.~I.}\ \bibnamefont
  {Kolobov}},\ }\bibfield  {title} {\bibinfo {title} {The spatial behavior of
  nonclassical light},\ }\href {https://doi.org/10.1103/RevModPhys.71.1539}
  {\bibfield  {journal} {\bibinfo  {journal} {Rev. Mod. Phys.}\ }\textbf
  {\bibinfo {volume} {71}},\ \bibinfo {pages} {1539} (\bibinfo {year}
  {1999})}\BibitemShut {NoStop}%
\bibitem [{\citenamefont {Corti}\ \emph {et~al.}(2016)\citenamefont {Corti},
  \citenamefont {Brambilla},\ and\ \citenamefont {Gatti}}]{Corti16}%
  \BibitemOpen
  \bibfield  {author} {\bibinfo {author} {\bibfnamefont {T.}~\bibnamefont
  {Corti}}, \bibinfo {author} {\bibfnamefont {E.}~\bibnamefont {Brambilla}},\
  and\ \bibinfo {author} {\bibfnamefont {A.}~\bibnamefont {Gatti}},\ }\bibfield
   {title} {\bibinfo {title} {Critical behavior of coherence and correlation of
  counterpropagating twin beams},\ }\href
  {https://doi.org/10.1103/PhysRevA.93.023837} {\bibfield  {journal} {\bibinfo
  {journal} {Phys. Rev. A}\ }\textbf {\bibinfo {volume} {93}},\ \bibinfo
  {pages} {023837} (\bibinfo {year} {2016})}\BibitemShut {NoStop}%
\bibitem [{\citenamefont {Horoshko}\ and\ \citenamefont
  {Kolobov}(2017)}]{Horoshko17}%
  \BibitemOpen
  \bibfield  {author} {\bibinfo {author} {\bibfnamefont {D.~B.}\ \bibnamefont
  {Horoshko}}\ and\ \bibinfo {author} {\bibfnamefont {M.~I.}\ \bibnamefont
  {Kolobov}},\ }\bibfield  {title} {\bibinfo {title} {Generation of monocycle
  squeezed light in chirped quasi-phase-matched nonlinear crystals},\ }\href
  {https://doi.org/10.1103/PhysRevA.95.033837} {\bibfield  {journal} {\bibinfo
  {journal} {Phys. Rev. A}\ }\textbf {\bibinfo {volume} {95}},\ \bibinfo
  {pages} {033837} (\bibinfo {year} {2017})}\BibitemShut {NoStop}%
\bibitem [{\citenamefont {Wasilewski}\ \emph {et~al.}(2006)\citenamefont
  {Wasilewski}, \citenamefont {Lvovsky}, \citenamefont {Banaszek},\ and\
  \citenamefont {Radzewicz}}]{Wasilewski06}%
  \BibitemOpen
  \bibfield  {author} {\bibinfo {author} {\bibfnamefont {W.}~\bibnamefont
  {Wasilewski}}, \bibinfo {author} {\bibfnamefont {A.~I.}\ \bibnamefont
  {Lvovsky}}, \bibinfo {author} {\bibfnamefont {K.}~\bibnamefont {Banaszek}},\
  and\ \bibinfo {author} {\bibfnamefont {C.}~\bibnamefont {Radzewicz}},\
  }\bibfield  {title} {\bibinfo {title} {Pulsed squeezed light: Simultaneous
  squeezing of multiple modes},\ }\href
  {https://doi.org/10.1103/PhysRevA.73.063819} {\bibfield  {journal} {\bibinfo
  {journal} {Phys. Rev. A}\ }\textbf {\bibinfo {volume} {73}},\ \bibinfo
  {pages} {063819} (\bibinfo {year} {2006})}\BibitemShut {NoStop}%
\bibitem [{\citenamefont {Quesada}\ \emph {et~al.}(2020)\citenamefont
  {Quesada}, \citenamefont {Triginer}, \citenamefont {Vidrighin},\ and\
  \citenamefont {Sipe}}]{Quesada20}%
  \BibitemOpen
  \bibfield  {author} {\bibinfo {author} {\bibfnamefont {N.}~\bibnamefont
  {Quesada}}, \bibinfo {author} {\bibfnamefont {G.}~\bibnamefont {Triginer}},
  \bibinfo {author} {\bibfnamefont {M.~D.}\ \bibnamefont {Vidrighin}},\ and\
  \bibinfo {author} {\bibfnamefont {J.~E.}\ \bibnamefont {Sipe}},\ }\bibfield
  {title} {\bibinfo {title} {Theory of high-gain twin-beam generation in
  waveguides: From {Maxwell's} equations to efficient simulation},\ }\href
  {https://doi.org/10.1103/PhysRevA.102.033519} {\bibfield  {journal} {\bibinfo
   {journal} {Phys. Rev. A}\ }\textbf {\bibinfo {volume} {102}},\ \bibinfo
  {pages} {033519} (\bibinfo {year} {2020})}\BibitemShut {NoStop}%
\bibitem [{\citenamefont {Horoshko}\ \emph {et~al.}(2019)\citenamefont
  {Horoshko}, \citenamefont {La~Volpe}, \citenamefont {Arzani}, \citenamefont
  {Treps}, \citenamefont {Fabre},\ and\ \citenamefont {Kolobov}}]{Horoshko19}%
  \BibitemOpen
  \bibfield  {author} {\bibinfo {author} {\bibfnamefont {D.~B.}\ \bibnamefont
  {Horoshko}}, \bibinfo {author} {\bibfnamefont {L.}~\bibnamefont {La~Volpe}},
  \bibinfo {author} {\bibfnamefont {F.}~\bibnamefont {Arzani}}, \bibinfo
  {author} {\bibfnamefont {N.}~\bibnamefont {Treps}}, \bibinfo {author}
  {\bibfnamefont {C.}~\bibnamefont {Fabre}},\ and\ \bibinfo {author}
  {\bibfnamefont {M.~I.}\ \bibnamefont {Kolobov}},\ }\bibfield  {title}
  {\bibinfo {title} {Bloch-{Messiah} reduction for twin beams of light},\
  }\href {https://doi.org/10.1103/PhysRevA.100.013837} {\bibfield  {journal}
  {\bibinfo  {journal} {Phys. Rev. A}\ }\textbf {\bibinfo {volume} {100}},\
  \bibinfo {pages} {013837} (\bibinfo {year} {2019})}\BibitemShut {NoStop}%
\bibitem [{\citenamefont {La~Volpe}\ \emph {et~al.}(2021)\citenamefont
  {La~Volpe}, \citenamefont {De}, \citenamefont {Kolobov}, \citenamefont
  {Parigi}, \citenamefont {Fabre}, \citenamefont {Treps},\ and\ \citenamefont
  {Horoshko}}]{LaVolpe21}%
  \BibitemOpen
  \bibfield  {author} {\bibinfo {author} {\bibfnamefont {L.}~\bibnamefont
  {La~Volpe}}, \bibinfo {author} {\bibfnamefont {S.}~\bibnamefont {De}},
  \bibinfo {author} {\bibfnamefont {M.~I.}\ \bibnamefont {Kolobov}}, \bibinfo
  {author} {\bibfnamefont {V.}~\bibnamefont {Parigi}}, \bibinfo {author}
  {\bibfnamefont {C.}~\bibnamefont {Fabre}}, \bibinfo {author} {\bibfnamefont
  {N.}~\bibnamefont {Treps}},\ and\ \bibinfo {author} {\bibfnamefont {D.~B.}\
  \bibnamefont {Horoshko}},\ }\bibfield  {title} {\bibinfo {title}
  {Spatiotemporal entanglement in a noncollinear optical parametric
  amplifier},\ }\href {https://doi.org/10.1103/PhysRevApplied.15.024016}
  {\bibfield  {journal} {\bibinfo  {journal} {Phys. Rev. Applied}\ }\textbf
  {\bibinfo {volume} {15}},\ \bibinfo {pages} {024016} (\bibinfo {year}
  {2021})}\BibitemShut {NoStop}%
\bibitem [{\citenamefont {Smirnova}\ \emph {et~al.}(2020)\citenamefont
  {Smirnova}, \citenamefont {Leykam}, \citenamefont {Chong},\ and\
  \citenamefont {Kivshar}}]{Smirnova20}%
  \BibitemOpen
  \bibfield  {author} {\bibinfo {author} {\bibfnamefont {D.}~\bibnamefont
  {Smirnova}}, \bibinfo {author} {\bibfnamefont {D.}~\bibnamefont {Leykam}},
  \bibinfo {author} {\bibfnamefont {Y.}~\bibnamefont {Chong}},\ and\ \bibinfo
  {author} {\bibfnamefont {Y.}~\bibnamefont {Kivshar}},\ }\bibfield  {title}
  {\bibinfo {title} {Nonlinear topological photonics},\ }\href
  {https://doi.org/10.1063/1.5142397} {\bibfield  {journal} {\bibinfo
  {journal} {Appl. Phys. Rev.}\ }\textbf {\bibinfo {volume} {7}},\ \bibinfo
  {pages} {021306} (\bibinfo {year} {2020})}\BibitemShut {NoStop}%
\bibitem [{\citenamefont {Barral}\ \emph
  {et~al.}(2020{\natexlab{a}})\citenamefont {Barral}, \citenamefont
  {Walschaers}, \citenamefont {Bencheikh}, \citenamefont {Parigi},
  \citenamefont {Levenson}, \citenamefont {Treps},\ and\ \citenamefont
  {Belabas}}]{Barral20a}%
  \BibitemOpen
  \bibfield  {author} {\bibinfo {author} {\bibfnamefont {D.}~\bibnamefont
  {Barral}}, \bibinfo {author} {\bibfnamefont {M.}~\bibnamefont {Walschaers}},
  \bibinfo {author} {\bibfnamefont {K.}~\bibnamefont {Bencheikh}}, \bibinfo
  {author} {\bibfnamefont {V.}~\bibnamefont {Parigi}}, \bibinfo {author}
  {\bibfnamefont {J.~A.}\ \bibnamefont {Levenson}}, \bibinfo {author}
  {\bibfnamefont {N.}~\bibnamefont {Treps}},\ and\ \bibinfo {author}
  {\bibfnamefont {N.}~\bibnamefont {Belabas}},\ }\bibfield  {title} {\bibinfo
  {title} {Quantum state engineering in arrays of nonlinear waveguides},\
  }\href {https://doi.org/10.1103/PhysRevA.102.043706} {\bibfield  {journal}
  {\bibinfo  {journal} {Phys. Rev. A}\ }\textbf {\bibinfo {volume} {102}},\
  \bibinfo {pages} {043706} (\bibinfo {year} {2020}{\natexlab{a}})}\BibitemShut
  {NoStop}%
\bibitem [{\citenamefont {Barral}\ \emph
  {et~al.}(2020{\natexlab{b}})\citenamefont {Barral}, \citenamefont
  {Walschaers}, \citenamefont {Bencheikh}, \citenamefont {Parigi},
  \citenamefont {Levenson}, \citenamefont {Treps},\ and\ \citenamefont
  {Belabas}}]{Barral20b}%
  \BibitemOpen
  \bibfield  {author} {\bibinfo {author} {\bibfnamefont {D.}~\bibnamefont
  {Barral}}, \bibinfo {author} {\bibfnamefont {M.}~\bibnamefont {Walschaers}},
  \bibinfo {author} {\bibfnamefont {K.}~\bibnamefont {Bencheikh}}, \bibinfo
  {author} {\bibfnamefont {V.}~\bibnamefont {Parigi}}, \bibinfo {author}
  {\bibfnamefont {J.~A.}\ \bibnamefont {Levenson}}, \bibinfo {author}
  {\bibfnamefont {N.}~\bibnamefont {Treps}},\ and\ \bibinfo {author}
  {\bibfnamefont {N.}~\bibnamefont {Belabas}},\ }\bibfield  {title} {\bibinfo
  {title} {Versatile photonic entanglement synthesizer in the spatial domain},\
  }\href {https://doi.org/10.1103/PhysRevApplied.14.044025} {\bibfield
  {journal} {\bibinfo  {journal} {Phys. Rev. Applied}\ }\textbf {\bibinfo
  {volume} {14}},\ \bibinfo {pages} {044025} (\bibinfo {year}
  {2020}{\natexlab{b}})}\BibitemShut {NoStop}%
\bibitem [{\citenamefont {Blanco-Redondo}\ \emph {et~al.}(2018)\citenamefont
  {Blanco-Redondo}, \citenamefont {Bell}, \citenamefont {Oren}, \citenamefont
  {Eggleton},\ and\ \citenamefont {Segev}}]{Blanco18}%
  \BibitemOpen
  \bibfield  {author} {\bibinfo {author} {\bibfnamefont {A.}~\bibnamefont
  {Blanco-Redondo}}, \bibinfo {author} {\bibfnamefont {B.}~\bibnamefont
  {Bell}}, \bibinfo {author} {\bibfnamefont {D.}~\bibnamefont {Oren}}, \bibinfo
  {author} {\bibfnamefont {B.~J.}\ \bibnamefont {Eggleton}},\ and\ \bibinfo
  {author} {\bibfnamefont {M.}~\bibnamefont {Segev}},\ }\bibfield  {title}
  {\bibinfo {title} {Topological protection of biphoton states},\ }\href@noop
  {} {\bibfield  {journal} {\bibinfo  {journal} {Science}\ }\textbf {\bibinfo
  {volume} {362}},\ \bibinfo {pages} {568} (\bibinfo {year}
  {2018})}\BibitemShut {NoStop}%
\bibitem [{\citenamefont {Christ}\ \emph {et~al.}(2013)\citenamefont {Christ},
  \citenamefont {Brecht}, \citenamefont {Mauerer},\ and\ \citenamefont
  {Silberhorn}}]{Christ13}%
  \BibitemOpen
  \bibfield  {author} {\bibinfo {author} {\bibfnamefont {A.}~\bibnamefont
  {Christ}}, \bibinfo {author} {\bibfnamefont {B.}~\bibnamefont {Brecht}},
  \bibinfo {author} {\bibfnamefont {W.}~\bibnamefont {Mauerer}},\ and\ \bibinfo
  {author} {\bibfnamefont {C.}~\bibnamefont {Silberhorn}},\ }\bibfield  {title}
  {\bibinfo {title} {Theory of quantum frequency conversion and {type-II}
  parametric down-conversion in the high-gain regime},\ }\href
  {http://stacks.iop.org/1367-2630/15/i=5/a=053038} {\bibfield  {journal}
  {\bibinfo  {journal} {New J. Phys.}\ }\textbf {\bibinfo {volume} {15}},\
  \bibinfo {pages} {053038} (\bibinfo {year} {2013})}\BibitemShut {NoStop}%
\bibitem [{\citenamefont {Quesada}\ and\ \citenamefont
  {Sipe}(2015)}]{Quesada15}%
  \BibitemOpen
  \bibfield  {author} {\bibinfo {author} {\bibfnamefont {N.}~\bibnamefont
  {Quesada}}\ and\ \bibinfo {author} {\bibfnamefont {J.~E.}\ \bibnamefont
  {Sipe}},\ }\bibfield  {title} {\bibinfo {title} {Time-ordering effects in the
  generation of entangled photons using nonlinear optical processes},\ }\href
  {https://doi.org/10.1103/PhysRevLett.114.093903} {\bibfield  {journal}
  {\bibinfo  {journal} {Phys. Rev. Lett.}\ }\textbf {\bibinfo {volume} {114}},\
  \bibinfo {pages} {093903} (\bibinfo {year} {2015})}\BibitemShut {NoStop}%
\bibitem [{\citenamefont {Lipfert}\ \emph {et~al.}(2018)\citenamefont
  {Lipfert}, \citenamefont {Horoshko}, \citenamefont {Patera},\ and\
  \citenamefont {Kolobov}}]{Lipfert18}%
  \BibitemOpen
  \bibfield  {author} {\bibinfo {author} {\bibfnamefont {T.}~\bibnamefont
  {Lipfert}}, \bibinfo {author} {\bibfnamefont {D.~B.}\ \bibnamefont
  {Horoshko}}, \bibinfo {author} {\bibfnamefont {G.}~\bibnamefont {Patera}},\
  and\ \bibinfo {author} {\bibfnamefont {M.~I.}\ \bibnamefont {Kolobov}},\
  }\bibfield  {title} {\bibinfo {title} {Bloch-{Messiah} decomposition and
  {Magnus} expansion for parametric down-conversion with monochromatic pump},\
  }\href {https://doi.org/10.1103/PhysRevA.98.013815} {\bibfield  {journal}
  {\bibinfo  {journal} {Phys. Rev. A}\ }\textbf {\bibinfo {volume} {98}},\
  \bibinfo {pages} {013815} (\bibinfo {year} {2018})}\BibitemShut {NoStop}%
\bibitem [{\citenamefont {Abram}(1987)}]{Abram87}%
  \BibitemOpen
  \bibfield  {author} {\bibinfo {author} {\bibfnamefont {I.}~\bibnamefont
  {Abram}},\ }\bibfield  {title} {\bibinfo {title} {Quantum theory of light
  propagation: {Linear} medium},\ }\href
  {https://doi.org/10.1103/PhysRevA.35.4661} {\bibfield  {journal} {\bibinfo
  {journal} {Phys. Rev. A}\ }\textbf {\bibinfo {volume} {35}},\ \bibinfo
  {pages} {4661} (\bibinfo {year} {1987})}\BibitemShut {NoStop}%
\bibitem [{\citenamefont {Huttner}\ \emph {et~al.}(1990)\citenamefont
  {Huttner}, \citenamefont {Serulnik},\ and\ \citenamefont
  {Ben-Aryeh}}]{Huttner90}%
  \BibitemOpen
  \bibfield  {author} {\bibinfo {author} {\bibfnamefont {B.}~\bibnamefont
  {Huttner}}, \bibinfo {author} {\bibfnamefont {S.}~\bibnamefont {Serulnik}},\
  and\ \bibinfo {author} {\bibfnamefont {Y.}~\bibnamefont {Ben-Aryeh}},\
  }\bibfield  {title} {\bibinfo {title} {Quantum analysis of light propagation
  in a parametric amplifier},\ }\href
  {https://doi.org/10.1103/PhysRevA.42.5594} {\bibfield  {journal} {\bibinfo
  {journal} {Phys. Rev. A}\ }\textbf {\bibinfo {volume} {42}},\ \bibinfo
  {pages} {5594} (\bibinfo {year} {1990})}\BibitemShut {NoStop}%
\bibitem [{\citenamefont {Serulnik}\ and\ \citenamefont
  {Ben-Aryeh}(1991)}]{Serulnik91}%
  \BibitemOpen
  \bibfield  {author} {\bibinfo {author} {\bibfnamefont {S.}~\bibnamefont
  {Serulnik}}\ and\ \bibinfo {author} {\bibfnamefont {Y.}~\bibnamefont
  {Ben-Aryeh}},\ }\bibfield  {title} {\bibinfo {title} {Space-time description
  of propagation in nonlinear dielectric media},\ }\href
  {https://doi.org/10.1088/0954-8998/3/1/006} {\bibfield  {journal} {\bibinfo
  {journal} {Quantum Opt. B}\ }\textbf {\bibinfo {volume} {3}},\ \bibinfo
  {pages} {63} (\bibinfo {year} {1991})}\BibitemShut {NoStop}%
\bibitem [{\citenamefont {Ben-Aryeh}\ and\ \citenamefont
  {Serulnik}(1991)}]{Ben-Aryeh91}%
  \BibitemOpen
  \bibfield  {author} {\bibinfo {author} {\bibfnamefont {Y.}~\bibnamefont
  {Ben-Aryeh}}\ and\ \bibinfo {author} {\bibfnamefont {S.}~\bibnamefont
  {Serulnik}},\ }\bibfield  {title} {\bibinfo {title} {The quantum treatment of
  propagation in non-linear optical media by the use of temporal modes},\
  }\href {https://doi.org/https://doi.org/10.1016/0375-9601(91)90650-W}
  {\bibfield  {journal} {\bibinfo  {journal} {Phys. Lett. A}\ }\textbf
  {\bibinfo {volume} {155}},\ \bibinfo {pages} {473} (\bibinfo {year}
  {1991})}\BibitemShut {NoStop}%
\bibitem [{\citenamefont {Toren}\ and\ \citenamefont
  {Ben-Aryeh}(1994)}]{Toren94}%
  \BibitemOpen
  \bibfield  {author} {\bibinfo {author} {\bibfnamefont {M.}~\bibnamefont
  {Toren}}\ and\ \bibinfo {author} {\bibfnamefont {Y.}~\bibnamefont
  {Ben-Aryeh}},\ }\bibfield  {title} {\bibinfo {title} {The problem of
  propagation in quantum optics, with applications to amplification, coupling
  of {EM} modes and distributed feedback lasers},\ }\href
  {https://doi.org/10.1088/0954-8998/6/5/006} {\bibfield  {journal} {\bibinfo
  {journal} {Quantum Opt. B}\ }\textbf {\bibinfo {volume} {6}},\ \bibinfo
  {pages} {425} (\bibinfo {year} {1994})}\BibitemShut {NoStop}%
\bibitem [{\citenamefont {Ben-Aryeh}\ \emph {et~al.}(1992)\citenamefont
  {Ben-Aryeh}, \citenamefont {Luk\v{s}},\ and\ \citenamefont
  {Pe\v{r}inov\'{a}}}]{Ben-Aryeh92}%
  \BibitemOpen
  \bibfield  {author} {\bibinfo {author} {\bibfnamefont {Y.}~\bibnamefont
  {Ben-Aryeh}}, \bibinfo {author} {\bibfnamefont {A.}~\bibnamefont
  {Luk\v{s}}},\ and\ \bibinfo {author} {\bibfnamefont {V.}~\bibnamefont
  {Pe\v{r}inov\'{a}}},\ }\bibfield  {title} {\bibinfo {title} {The concept of
  equal space commutators in quantum optics},\ }\href
  {https://doi.org/https://doi.org/10.1016/0375-9601(92)91047-U} {\bibfield
  {journal} {\bibinfo  {journal} {Phys. Lett. A}\ }\textbf {\bibinfo {volume}
  {165}},\ \bibinfo {pages} {19} (\bibinfo {year} {1992})}\BibitemShut
  {NoStop}%
\bibitem [{\citenamefont {Pe{\v{r}}ina}\ and\ \citenamefont {{Pe{\v{r}}ina,
  Jr.}}(1995)}]{Perina95}%
  \BibitemOpen
  \bibfield  {author} {\bibinfo {author} {\bibfnamefont {J.}~\bibnamefont
  {Pe{\v{r}}ina}}\ and\ \bibinfo {author} {\bibfnamefont {J.}~\bibnamefont
  {{Pe{\v{r}}ina, Jr.}}},\ }\bibfield  {title} {\bibinfo {title} {Photon
  statistics of a contradirectional nonlinear coupler},\ }\href
  {https://doi.org/10.1088/1355-5111/7/5/007} {\bibfield  {journal} {\bibinfo
  {journal} {Quantum Semiclass. Opt. B}\ }\textbf {\bibinfo {volume} {7}},\
  \bibinfo {pages} {849} (\bibinfo {year} {1995})}\BibitemShut {NoStop}%
\bibitem [{\citenamefont {{Pe{\v{r}}ina, Jr}}\ and\ \citenamefont
  {Pe{\v{r}}ina}(2000)}]{Perina00}%
  \BibitemOpen
  \bibfield  {author} {\bibinfo {author} {\bibfnamefont {J.}~\bibnamefont
  {{Pe{\v{r}}ina, Jr}}}\ and\ \bibinfo {author} {\bibfnamefont
  {J.}~\bibnamefont {Pe{\v{r}}ina}},\ }\bibfield  {title} {\bibinfo {title}
  {Quantum statistics of nonlinear optical couplers},\ }\href@noop {}
  {\bibfield  {journal} {\bibinfo  {journal} {Prog. Opt.}\ }\textbf {\bibinfo
  {volume} {41}},\ \bibinfo {pages} {361} (\bibinfo {year} {2000})}\BibitemShut
  {NoStop}%
\bibitem [{\citenamefont {Luk{\v{s}}}\ and\ \citenamefont
  {Pe\v{r}inova}(2002)}]{Luks02}%
  \BibitemOpen
  \bibfield  {author} {\bibinfo {author} {\bibfnamefont {A.}~\bibnamefont
  {Luk{\v{s}}}}\ and\ \bibinfo {author} {\bibfnamefont {V.}~\bibnamefont
  {Pe\v{r}inova}},\ }\bibfield  {title} {\bibinfo {title} {Canonical quantum
  description of light propagation in dielectric media},\ }\href@noop {}
  {\bibfield  {journal} {\bibinfo  {journal} {Prog. Opt.}\ }\textbf {\bibinfo
  {volume} {43}},\ \bibinfo {pages} {295} (\bibinfo {year} {2002})}\BibitemShut
  {NoStop}%
\bibitem [{\citenamefont {Luk{\v{s}}}\ and\ \citenamefont
  {Pe\v{r}inova}(2009)}]{Luks09}%
  \BibitemOpen
  \bibfield  {author} {\bibinfo {author} {\bibfnamefont {A.}~\bibnamefont
  {Luk{\v{s}}}}\ and\ \bibinfo {author} {\bibfnamefont {V.}~\bibnamefont
  {Pe\v{r}inova}},\ }\href@noop {} {\emph {\bibinfo {title} {Quantum Aspects of
  Light Propagation}}}\ (\bibinfo  {publisher} {Springer},\ \bibinfo {address}
  {New York},\ \bibinfo {year} {2009})\BibitemShut {NoStop}%
\bibitem [{\citenamefont {Novikov}\ and\ \citenamefont
  {Novikov}(2013)}]{Novikov-Novikov}%
  \BibitemOpen
  \bibfield  {author} {\bibinfo {author} {\bibfnamefont {A.~M.}\ \bibnamefont
  {Novikov}}\ and\ \bibinfo {author} {\bibfnamefont {D.~A.}\ \bibnamefont
  {Novikov}},\ }\href {https://doi.org/10.1201/b14562} {\emph {\bibinfo {title}
  {Research methodology: From philosophy of science to research design}}}\
  (\bibinfo  {publisher} {CRC Press},\ \bibinfo {address} {London},\ \bibinfo
  {year} {2013})\BibitemShut {NoStop}%
\bibitem [{\citenamefont {Dirac}(1964)}]{Dirac64}%
  \BibitemOpen
  \bibfield  {author} {\bibinfo {author} {\bibfnamefont {P.~A.~M.}\
  \bibnamefont {Dirac}},\ }\href@noop {} {\emph {\bibinfo {title} {Lectures on
  Quantum Mechanics}}}\ (\bibinfo  {publisher} {Yeshiva University},\ \bibinfo
  {address} {New York},\ \bibinfo {year} {1964})\BibitemShut {NoStop}%
\bibitem [{\citenamefont {Akhiezer}\ and\ \citenamefont
  {Berestetsky}(1965)}]{Akhiezer65}%
  \BibitemOpen
  \bibfield  {author} {\bibinfo {author} {\bibfnamefont {A.~I.}\ \bibnamefont
  {Akhiezer}}\ and\ \bibinfo {author} {\bibfnamefont {V.~B.}\ \bibnamefont
  {Berestetsky}},\ }\href@noop {} {\emph {\bibinfo {title} {Quantum
  Electrodynamics}}}\ (\bibinfo  {publisher} {Interscience Publishers},\
  \bibinfo {address} {New York},\ \bibinfo {year} {1965})\BibitemShut {NoStop}%
\bibitem [{\citenamefont {Goldstein}(1980)}]{Goldstein80}%
  \BibitemOpen
  \bibfield  {author} {\bibinfo {author} {\bibfnamefont {H.}~\bibnamefont
  {Goldstein}},\ }\href@noop {} {\emph {\bibinfo {title} {Classical
  Mechanics}}},\ \bibinfo {edition} {2nd}\ ed.\ (\bibinfo  {publisher} {Addison
  Wesley},\ \bibinfo {address} {Reading},\ \bibinfo {year} {1980})\BibitemShut
  {NoStop}%
\bibitem [{\citenamefont {Landau}\ and\ \citenamefont
  {Lifshitz}(1981)}]{Landau-LifshitzI}%
  \BibitemOpen
  \bibfield  {author} {\bibinfo {author} {\bibfnamefont {L.~D.}\ \bibnamefont
  {Landau}}\ and\ \bibinfo {author} {\bibfnamefont {E.~M.}\ \bibnamefont
  {Lifshitz}},\ }\href@noop {} {\emph {\bibinfo {title} {Mechanics}}},\
  \bibinfo {edition} {3rd}\ ed.\ (\bibinfo  {publisher}
  {Butterworth-Heinemann},\ \bibinfo {address} {Oxford},\ \bibinfo {year}
  {1981})\BibitemShut {NoStop}%
\bibitem [{\citenamefont {Glauber}(1963{\natexlab{a}})}]{Glauber63b}%
  \BibitemOpen
  \bibfield  {author} {\bibinfo {author} {\bibfnamefont {R.~J.}\ \bibnamefont
  {Glauber}},\ }\bibfield  {title} {\bibinfo {title} {Coherent and incoherent
  states of the radiation field},\ }\href
  {https://doi.org/10.1103/PhysRev.131.2766} {\bibfield  {journal} {\bibinfo
  {journal} {Phys. Rev.}\ }\textbf {\bibinfo {volume} {131}},\ \bibinfo {pages}
  {2766} (\bibinfo {year} {1963}{\natexlab{a}})}\BibitemShut {NoStop}%
\bibitem [{\citenamefont {Mandel}\ and\ \citenamefont
  {Wolf}(1995)}]{Mandel-Wolf}%
  \BibitemOpen
  \bibfield  {author} {\bibinfo {author} {\bibfnamefont {L.}~\bibnamefont
  {Mandel}}\ and\ \bibinfo {author} {\bibfnamefont {E.}~\bibnamefont {Wolf}},\
  }\href@noop {} {\emph {\bibinfo {title} {Optical coherence and quantum
  optics}}}\ (\bibinfo  {publisher} {Cambridge University},\ \bibinfo {address}
  {Cambridge},\ \bibinfo {year} {1995})\BibitemShut {NoStop}%
\bibitem [{\citenamefont {Glauber}(1963{\natexlab{b}})}]{Glauber63a}%
  \BibitemOpen
  \bibfield  {author} {\bibinfo {author} {\bibfnamefont {R.~J.}\ \bibnamefont
  {Glauber}},\ }\bibfield  {title} {\bibinfo {title} {The quantum theory of
  optical coherence},\ }\href {https://doi.org/10.1103/PhysRev.130.2529}
  {\bibfield  {journal} {\bibinfo  {journal} {Phys. Rev.}\ }\textbf {\bibinfo
  {volume} {130}},\ \bibinfo {pages} {2529} (\bibinfo {year}
  {1963}{\natexlab{b}})}\BibitemShut {NoStop}%
\bibitem [{\citenamefont {Landau}\ and\ \citenamefont
  {Lifshitz}(1987)}]{Landau-LifshitzII}%
  \BibitemOpen
  \bibfield  {author} {\bibinfo {author} {\bibfnamefont {L.~D.}\ \bibnamefont
  {Landau}}\ and\ \bibinfo {author} {\bibfnamefont {E.~M.}\ \bibnamefont
  {Lifshitz}},\ }\href@noop {} {\emph {\bibinfo {title} {The Classical Theory
  of Fields}}},\ \bibinfo {edition} {4th}\ ed.\ (\bibinfo  {publisher}
  {Butterworth-Heinemann},\ \bibinfo {address} {Oxford},\ \bibinfo {year}
  {1987})\BibitemShut {NoStop}%
\bibitem [{\citenamefont {Jackson}(1999)}]{Jackson}%
  \BibitemOpen
  \bibfield  {author} {\bibinfo {author} {\bibfnamefont {J.~D.}\ \bibnamefont
  {Jackson}},\ }\href@noop {} {\emph {\bibinfo {title} {Classical
  Electrodynamics}}},\ \bibinfo {edition} {3rd}\ ed.\ (\bibinfo  {publisher}
  {Wiley},\ \bibinfo {address} {New York},\ \bibinfo {year} {1999})\BibitemShut
  {NoStop}%
\bibitem [{\citenamefont {Dirac}(1958)}]{Dirac58}%
  \BibitemOpen
  \bibfield  {author} {\bibinfo {author} {\bibfnamefont {P.~A.~M.}\
  \bibnamefont {Dirac}},\ }\href@noop {} {\emph {\bibinfo {title} {The
  Principles of Quantum Mechanics}}},\ \bibinfo {edition} {4th}\ ed.\ (\bibinfo
   {publisher} {Oxford University},\ \bibinfo {address} {Oxford},\ \bibinfo
  {year} {1958})\BibitemShut {NoStop}%
\bibitem [{\citenamefont {Jackson}(2002)}]{Jackson02}%
  \BibitemOpen
  \bibfield  {author} {\bibinfo {author} {\bibfnamefont {J.~D.}\ \bibnamefont
  {Jackson}},\ }\bibfield  {title} {\bibinfo {title} {From {Lorenz to Coulomb}
  and other explicit gauge transformations},\ }\href
  {https://doi.org/10.1119/1.1491265} {\bibfield  {journal} {\bibinfo
  {journal} {Am. J. Phys.}\ }\textbf {\bibinfo {volume} {70}},\ \bibinfo
  {pages} {917} (\bibinfo {year} {2002})}\BibitemShut {NoStop}%
\bibitem [{\citenamefont {Horoshko}\ \emph {et~al.}(2018)\citenamefont
  {Horoshko}, \citenamefont {Eskandary},\ and\ \citenamefont
  {Kilin}}]{Horoshko18JOSAB}%
  \BibitemOpen
  \bibfield  {author} {\bibinfo {author} {\bibfnamefont {D.~B.}\ \bibnamefont
  {Horoshko}}, \bibinfo {author} {\bibfnamefont {M.~M.}\ \bibnamefont
  {Eskandary}},\ and\ \bibinfo {author} {\bibfnamefont {S.~Y.}\ \bibnamefont
  {Kilin}},\ }\bibfield  {title} {\bibinfo {title} {Quantum model for
  traveling-wave electro-optical phase modulator},\ }\href@noop {} {\bibfield
  {journal} {\bibinfo  {journal} {J. Opt. Soc. Am. B}\ }\textbf {\bibinfo
  {volume} {35}},\ \bibinfo {pages} {2744} (\bibinfo {year}
  {2018})}\BibitemShut {NoStop}%
\bibitem [{\citenamefont {Li{\~n}ares}\ \emph {et~al.}(2008)\citenamefont
  {Li{\~n}ares}, \citenamefont {Nistal},\ and\ \citenamefont
  {Barral}}]{Linares08}%
  \BibitemOpen
  \bibfield  {author} {\bibinfo {author} {\bibfnamefont {J.}~\bibnamefont
  {Li{\~n}ares}}, \bibinfo {author} {\bibfnamefont {M.}~\bibnamefont
  {Nistal}},\ and\ \bibinfo {author} {\bibfnamefont {D.}~\bibnamefont
  {Barral}},\ }\bibfield  {title} {\bibinfo {title} {Quantization of coupled
  {1D} vector modes in integrated photonic waveguides},\ }\href@noop {}
  {\bibfield  {journal} {\bibinfo  {journal} {New J. Phys.}\ }\textbf {\bibinfo
  {volume} {10}},\ \bibinfo {pages} {063023} (\bibinfo {year}
  {2008})}\BibitemShut {NoStop}%
\bibitem [{\citenamefont {Horoshko}\ \emph {et~al.}(2012)\citenamefont
  {Horoshko}, \citenamefont {Patera}, \citenamefont {Gatti},\ and\
  \citenamefont {Kolobov}}]{Horoshko12epjd}%
  \BibitemOpen
  \bibfield  {author} {\bibinfo {author} {\bibfnamefont {D.~B.}\ \bibnamefont
  {Horoshko}}, \bibinfo {author} {\bibfnamefont {G.}~\bibnamefont {Patera}},
  \bibinfo {author} {\bibfnamefont {A.}~\bibnamefont {Gatti}},\ and\ \bibinfo
  {author} {\bibfnamefont {M.~I.}\ \bibnamefont {Kolobov}},\ }\bibfield
  {title} {\bibinfo {title} {X-entangled biphotons: {Schmidt} number for {2D}
  model},\ }\href@noop {} {\bibfield  {journal} {\bibinfo  {journal} {Eur.
  Phys. J. D}\ }\textbf {\bibinfo {volume} {66}},\ \bibinfo {pages} {1}
  (\bibinfo {year} {2012})}\BibitemShut {NoStop}%
\bibitem [{\citenamefont {Barnett}\ and\ \citenamefont
  {Radmore}(1997)}]{Barnett97}%
  \BibitemOpen
  \bibfield  {author} {\bibinfo {author} {\bibfnamefont {S.~M.}\ \bibnamefont
  {Barnett}}\ and\ \bibinfo {author} {\bibfnamefont {P.~M.}\ \bibnamefont
  {Radmore}},\ }\href@noop {} {\emph {\bibinfo {title} {Methods in theoretical
  quantum optics}}}\ (\bibinfo  {publisher} {Clarendon Press},\ \bibinfo
  {address} {Oxford},\ \bibinfo {year} {1997})\BibitemShut {NoStop}%
\end{thebibliography}%
\end{document}